\documentclass[sigconf]{acmart}
\usepackage{multirow}
\usepackage{arydshln}
\usepackage{algorithmic}
\usepackage{algorithm}
\usepackage{subcaption}
\usepackage{mwe}
\usepackage{graphicx}
\usepackage{comment}
\usepackage{booktabs} 
\usepackage{colortbl}
\usepackage{enumitem}
\AtBeginDocument{% 
}

\copyrightyear{2025}
\acmYear{2025}
\setcopyright{rightsretained}
\acmConference[WSDM '25]{Proceedings of the Eighteenth ACM International Conference on Web Search and Data Mining}{March 10--14, 2025}{Hannover, Germany}
\acmBooktitle{Proceedings of the Eighteenth ACM International Conference on Web Search and Data Mining (WSDM '25), March 10--14, 2025, Hannover, Germany}
\acmDOI{10.1145/3701551.3703496}
\acmISBN{979-8-4007-1329-3/25/03}

\begin{document}


\title{Large Language Model driven Policy Exploration for Recommender Systems
}

\author{Jie Wang}
\orcid{0009-0005-7117-8580}
\email{j.wang.9@research.gla.ac.uk}
\affiliation{%
  \institution{University of Glasgow}
    \city{Glasgow}
  \country{UK}
}

\author{Alexandros Karatzoglou}
\orcid{0000-0001-6063-9023}
\email{alexandros.karatzoglou@gmail.com}
\affiliation{%
  \institution{Amazon}
    \city{Barcelona}
  \country{Spain}
}

\author{Ioannis Arapakis}
\orcid{0000-0003-2528-6597}
\email{ioannis.arapakis@telefonica.com}
\affiliation{%
  \institution{Telefonica Research}
  \city{Barcelona}
  \country{Spain}
}

\author{Joemon M. Jose}
\orcid{0000-0001-9228-1759}
\email{joemon.jose@glasgow.ac.uk}
\affiliation{%
 \institution{University of Glasgow}
   \city{Glasgow}
 \country{UK}
}

\renewcommand{\shortauthors}{Jie Wang, Alexandros Karatzoglou, Ioannis Arapakis, and Joemon M. Jose}

\begin{abstract}
Recent advancements in Recommender Systems (RS) have incorporated Reinforcement Learning (RL), framing the recommendation as a Markov Decision Process (MDP). However, offline RL policies trained on static user data are vulnerable to distribution shift when deployed in dynamic online environments. Additionally, excessive focus on exploiting short-term relevant items can hinder exploration, leading to sub-optimal recommendations and negatively impacting long-term user gains. Online RL-based RS also face challenges in production deployment, due to the risks of exposing users to untrained or unstable policies. Large Language Models (LLMs) offer a promising solution to mimic user objectives and preferences %through language
%, making them suitable 
for pre-training policies offline to enhance the initial recommendations in online settings. Effectively managing distribution shift and balancing exploration are crucial for improving RL-based RS, especially when leveraging LLM-based pre-training.

To address these challenges, we propose an Interaction-Augmented Learned Policy (iALP) that utilizes user preferences distilled from an LLM. Our approach involves prompting the LLM with user states to extract item preferences, learning rewards based on feedback, and updating the RL policy using an actor-critic framework. Furthermore, to deploy iALP in an online scenario, we introduce an adaptive variant, A-iALP, that implements a simple fine-tuning strategy (A-iALP$_{ft}$), and an adaptive approach (A-iALP$_{ap}$) designed to mitigate issues with compromised policies and limited exploration. Experiments across three simulated environments demonstrate that A-iALP introduces substantial performance improvements.

\end{abstract}

\begin{CCSXML}
<ccs2012>
 <concept>
  <concept_id>10010520.10010553.10010562</concept_id>
  <concept_desc>Computer systems organization~Embedded systems</concept_desc>
  <concept_significance>500</concept_significance>
 </concept>
 <concept>
  <concept_id>10010520.10010575.10010755</concept_id>
  <concept_desc>Computer systems organization~Redundancy</concept_desc>
  <concept_significance>300</concept_significance>
 </concept>
 <concept>
  <concept_id>10010520.10010553.10010554</concept_id>
  <concept_desc>Computer systems organization~Robotics</concept_desc>
  <concept_significance>100</concept_significance>
 </concept>
 <concept>
  <concept_id>10003033.10003083.10003095</concept_id>
  <concept_desc>Networks~Network reliability</concept_desc>
  <concept_significance>100</concept_significance>
 </concept>
</ccs2012>
\end{CCSXML}

%\ccsdesc[500]{Computer systems organization~Embedded systems}
\ccsdesc[500]{Information systems~Recommender systems}
%\ccsdesc[500]{Diversity and Novelty in recommendation}

\keywords{Online Reinforcement Learning for Recommendation; LLM Policy; Long-term Gains; Exploration Strategies }

 % \caption{Seattle Mariners at Spring Training, 2010.}
 % \Description{Enjoying the baseball game from the third-base

% \received{20 January 2023}

\maketitle

\section{Introduction}

Modelling user states through interacted items to generate future recommendations has become a common paradigm for many re-commender systems (RS), particularly in session-based or sequential models~\cite{hidasi2015session,tang2018personalized,yuan2019simple,kang2018self}. Traditionally, these methods implement offline training on historical user log data, using self-supervised techniques. To address long-term user benefits, recent research reformulated this process as a Markov Decision Process (MDP)~\cite{xin2020self,xin2022supervised,ren2023contrastive}, introducing Reinforcement Learning (RL) to solve the associated decision-making challenges.

However, efforts to develop offline RL policies using historical user data \cite{xin2020self,xin2022supervised,ren2023contrastive} have been shown to be vulnerable to distribution drift, when deployed online~\cite{farahmand2010,Kumar2019}. This can lead to %unexpected and 
erroneous actions in out-of-distribution states due to the absence of real-time user feedback from interactions with the environment. Furthermore, offline RL-based systems often suffer from insufficient exploration of the data space, resulting in sub-optimal model performance. These limitations are particularly pronounced in cold-start scenarios and during initial stages when training data is sparse, causing policies to perform poorly and fail to provide satisfactory recommendations. 
% Therefore, effective online methods are needed to improve long-term gains.
To address these challenges and improve long-term gains, effective methods that can thoroughly explore the data space while adapting to real-time user interactions are needed.

To this end, researchers have explored online RL methods for recommendation that learn from real-time user feedback, as illustrated in Figure~\ref{figure0}. For example, DQN~\cite{mnih2013playing}, A2C~\cite{konda1999actor}, and DDPG~\cite{lillicrap2015continuous} have been used to to optimize cumulative rewards, aiming to enhance long-term user engagement. However, these methods typically require extensive training iterations before achieving optimal performance.
Moreover, they overlook a crucial issue that can lead to user churn: if users receive unsatisfactory recommendations during their initial interactions with a system, they are likely to abandon it rather than stay and provide feedback for future improvements. This challenge is one of the main reasons~\cite{xin2020self} why online RL has not gained widespread adoption in RS.

LLMs with knowledge transfer capabilities have recently
gained significant attention in RS~\cite{he2023large,zhang2023user,du2023enhancing}.
In offline settings, LLMs have demonstrated potential as zero-shot~\cite{bao2023tallrec} or pre-trained~\cite{geng2022recommendation} RS, achieved through fine-tuning on user data. Moreover, LLMs have been shown to capture user objectives, learnt from expert demonstrations or preferences, through their intuitive use of language~\cite{wang2024reinforcement}. This suggests the potential to leverage LLMs to interpret user preferences and generate appropriate actions across various states. Such feedback can be utilised to pre-train a policy offline, enabling desirable initial behaviours for online recommendations.

To further enhance the performance of an RL-based RS, we introduce an Interaction-Augmented Learned Policy (iALP). We initially train the policy by exclusively interacting with an LLM. We first derive user preferences from an offline LLM, which are subsequently employed for online user interactions. To derive preferences for various items, we prompt the LLM with task-specific instructions to select potential actions. The LLM assigns rewards (1 or 0) for candidate items based on the prompted preferences. This process generates an interactive offline dataset, entirely generated by the LLM, which is used to continuously update the RL policy using an actor-critic framework~\cite{konda1999actor}.

For transitioning the policy to online, we propose an adaptive Interaction-Augmented Learned Policy (A-iALP) variant. 
Specifically, we introduce the direct fine-tuning strategy A-iALP$_{ft}$ to initialise and update the online agent with real feedback from the user environment. A-iALP$_{ft}$ can be deployed as an on-line agent to effectively address policy drift issues and limited exploration, two challenges with off-line RL. In addition, we propose a second adaptive approach, A-iALT$_{ap}$, which combines a frozen pretrained agent with an online policy to allow further learning. Specifically, the useful behaviours of iAPT are retained to generate desirable sequences in the initial training stage, and in the later stage the actions from the online policy are mainly used to interact with the environment. Furthermore, we validate the performance of A-iALP under different exploration strategies against established baselines. Experimental results on three simulated environments demonstrate the superior performance A-iAPT in generating desirable initial and stable episodes.
Our contributions can be summarised as follows:
 \begin{itemize}[leftmargin=*]

    \item We propose to distil user preferences  on items using LLMs and use these preferences to train an interaction-Augmented Learned Policy (iALP). 
   
    \item We introduce two adaptive methods for iALP-to-Online (A-iALP) RL-based recommendation, leading to faster and stable online policy convergence, to enhance long-term rewards (return) and alleviate gain loss for users at early steps. 
    
    \item  We conduct experiments on three recommendation datasets. Experimental results show significant improvements in all metrics.
 \end{itemize}

\section{Related Work}
%\section{Preliminary}
\subsection{RL for Recommendation} 
Recommenders,
e.g., session-based or sequential methods~\cite{hidasi2015session,tang2018personalized,yuan2019simple,kang2018self}, are often trained offline on historical user data in a self-supervised manner. 
To target long-term gains for users, recent research has reformulated the process as a Markov Decision Process (MDP), introducing RL to solve the associated decision-making challenges~\cite{zhao2018recommendations,wang2020kerl,wang2024sparks}. Specifically, ~\citet{liu2023exploration} extended DDPG to session-based recommendation, while
~\citet{xin2020self} introduced self-supervised RL for Sequential Q-Network (SQN) and Soft Actor-Critic (SAC) to improve the accuracy of the baselines with respect to user clicks and purchases.~\citet{xin2022supervised} further enhanced the frameworks with SNQN and SA2C by using a sampling strategy to integrate negative feedback. Another approach~\cite{xin2022rethinking} involves guiding training by modelling desirable cumulative rewards rather than focusing on expected returns at each timestamp. Furthermore, ~\citet{ren2023contrastive} increased the original states to improve SQN with contrastive learning. Despite these advancements, the aforementioned methods are primarily trained offline on historical log data, resulting to sub-optimal policies due to potential distribution drift when deployed online. Furthermore, focusing on recommending the most relevant items based on historical sessions can introduce biases that negatively impact user experience by neglecting long-term engagement. To optimise long-term gains for users, some recent work explored online RL-based recommendation and reward-related metrics~\cite{yu2024easyrl4rec}. %In our paper, we explore how to 
However, this approach requires extensive training steps to achieve satisfactory results. In our paper, we primarily explore how to accelerate the enhancement of users' long-term benefits while mitigating the loss of user satisfaction in the initial stage.

\subsection{LLMs for Recommendation}
LLMs ~\cite{zhao2023survey,liu2023pre,min2023recent,raffel2020exploring,touvron2023llama} pre-trained on massive natural language datasets with continuously enhanced transfer capabilities have gained attention in the field of RS~\cite{chen2023large,lin2023can,fan2023recommender, fu2024exploring}. Existing offline adaptations of LLMs for recommendation tasks involve primarily training the LLM to be a new recommender through pre-training~\cite{geng2022recommendation,hou2022towards,wang2022transrec}, fine-tuning~\cite{min2023recent}, prompt-based tuning~\cite{liu2023pre,ouyang2022training,bao2023tallrec} and item representation~\cite{rajput2023recommender}, data augmentation~\cite{wang2024reinforcement} etc.
For online RL-based recommendation, ~\cite{shi2024large} adapts LLM directly as an online agent in aspects of actor/critic/planner to improve long-term engagement of users.
However, these methods of employing LLMs as recommenders come with significant pre-training costs or face difficulties in preserving LLM signals.  
Recent work by~\citet{kwon2023reward} demonstrated that the use of LLMs as reward models can significantly outperform traditional learnt rewards in the context of RL applications for games. Inspired by this, we propose the utilisation of LLMs for specifying user objectives and preferences through natural language, making them suitable for pre-training policy offline to boot the initial recommendations online. In addition, such an approach facilitates rigourous exploration of data spaces to create effective models.

\begin{figure}[!t]
      \centering
        \includegraphics[trim={0 0 0 0}, clip, scale=0.26]{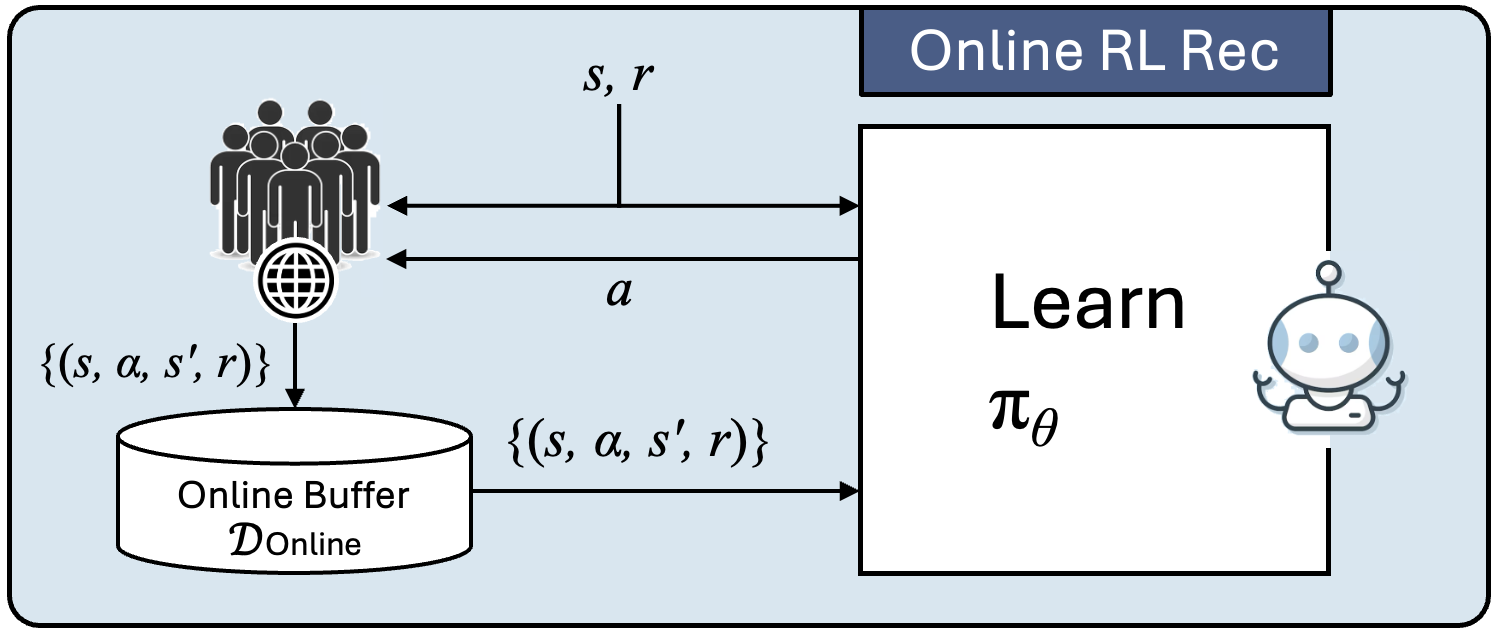}
    \vspace{-0.5em}
    \caption{
    Process of online RL for recommendation. %\textcolor{red}{Fig 1 not described in the text}
    }
    \label{figure0}
\end{figure}

\begin{table*}[] 
%\fontsize{9}{10}\selectfont
\setlength{\tabcolsep}{2.5pt}
\caption{Prompt template to generate user preference from LLM. 
}
\label{prompt}
\centering
\begin{tabular}{rp{14.5cm}}
\toprule
& Below is an instruction that describes a task, paired with an input that provides further context. Write a response that appropriately completes the request. \\ 
Instruction:      &You are a user in a \{\textbf{scenario}\} platform now. The \{\textbf{item}\} is in the form of \{\textbf{attribute1; attribute2; ... }\}.
Given a user's \{\textbf{behavior}\} history of \{\textbf{item}\}, and candidate \{\textbf{item}\} labelled by lowercase letter to be decided to recommend to the user, identify which \{\textbf{item}\} the user will mostly prefer to at next timestamp. 
Please judge by the user's preference on \{\textbf{attribute1; attribute2; ... }\}; if you think that none of the candidates will be selected by the user, please answer "None"       \\
Input:   & History: \{\textbf{interacted items}\}.
Which one the user will mostly like at next timestamp in the following candidates?
\newline a. \{\textbf{item1}\}
b. \{\textbf{item2}\}
...
j. \{\textbf{item10}\}
k. None         \\
Response:    &By analysing the user's preference, the user will select \{\textbf{label from a-k}\}  \\
\bottomrule
\end{tabular}
\end{table*}

\section{Preliminary}%

\subsection{RL-based Long-term Recommendation }
\label{task_formulation}
Online recommender systems aim to retrieve items that can enhance the user experience and provide long-term engagement.
RL-based RS for long-term user engagement/satisfaction is based on the principles of a Markov Decision Process (MDP)~\cite{xin2020self}. 
The agent (recommender) interacts with the environment (users), taking
actions (recommending items) based on the state of the environment, which is represented by the interacted items of the user. The agent is then updated by the feedback (reward) from the user. Continually, the environment generates a new state for the agent as users respond to the actions. Figure~\ref{figure0} shows the process, represented by the tuple $(\mathcal{S}, \mathcal{A}, \mathcal{P}, r, \gamma )$:

 \begin{itemize}[leftmargin=*]
    
     \item State space $\mathcal{S}$ represents the user's state. $s_{t} \in \mathcal{S}$ is the state at timestamp $t$, which is usually generated by mapping the sequence of interacted items (user history) into a sequential model/state encoder.
    
     \item Action space $\mathcal{A}$. The discrete action set is comprised of the candidate items. Taking action means recommending items. In the online setting, the action $a_t\in \mathcal{A}$ at timestamp $t$ will be the interacted item in the next timestamp to construct the next state.
     
    \item Reward function $r$ returns immediate reward $r_t$ given state $s_t$ and the action $a_{t}$ taken by the agent at timestamp $t$, which reflects the user's feedback on current recommendation, e.g., likes, dislikes.  

    \item State transition function $\mathcal{P}$ describes the next state  $s_{t+1}$ from the environment given the current state  $s_t$ and observed action $a_t$. 
     
    \item Discount factor $\gamma $ to the future rewards, where $\gamma  \in [0,1]$.
     
\end{itemize}

The RL-based recommendation aims to learn a target policy $\pi_\psi(a | s)$ that maps the state  $s\in\mathcal{S}$ to an action distribution $a\in\mathcal{A}$ by maximizing the expected cumulative rewards (return) to realize long-term user engagement, where $\theta$ denotes the parameters:
\begin{equation}
\max _{\pi_\theta} \mathbb{E}_{\tau \sim \pi_\theta}[R(\tau)] \text {, where } R(\tau)=\sum_{t=0}^{|\tau|} \gamma^t r\left(\mathrm{~s}_t, a_t\right) \text {, }
\end{equation}
where $\tau$ denotes the trajectory of $(s_t, a_t, s_{t+1})$. 
\begin{figure}[!t]
      \centering
        \includegraphics[trim={0 0 0 0}, clip, scale=0.14]{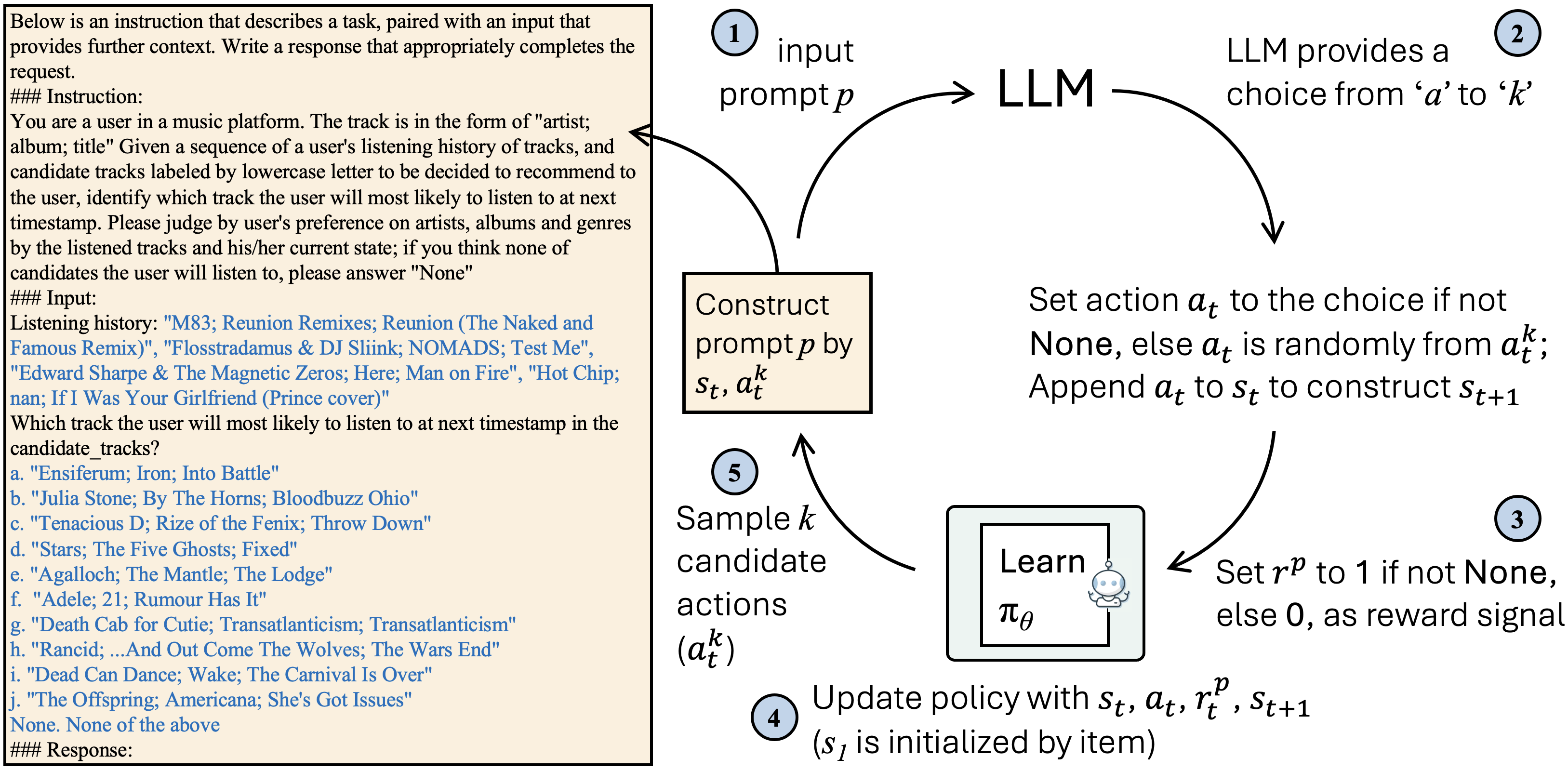}
    %\vspace{-0.5em}
    \caption{
    Generate user preference, e.g., action and reward, for offline pre-training.
    }
    \label{figure_preference}
\end{figure}

\begin{figure*}[!t]
    \captionsetup[subfloat]%{captionskip=-5pt,nearskip=0pt,farskip=0pt}
    {}
    \centering
    \begin{minipage}[c]{0.78\textwidth}
    \centering
        \subfloat[A-iALP$_{ft}$: Direct fine-tuning iALP $\theta$]{%
        \includegraphics[trim={0 340 0 0}, clip, scale=0.26]{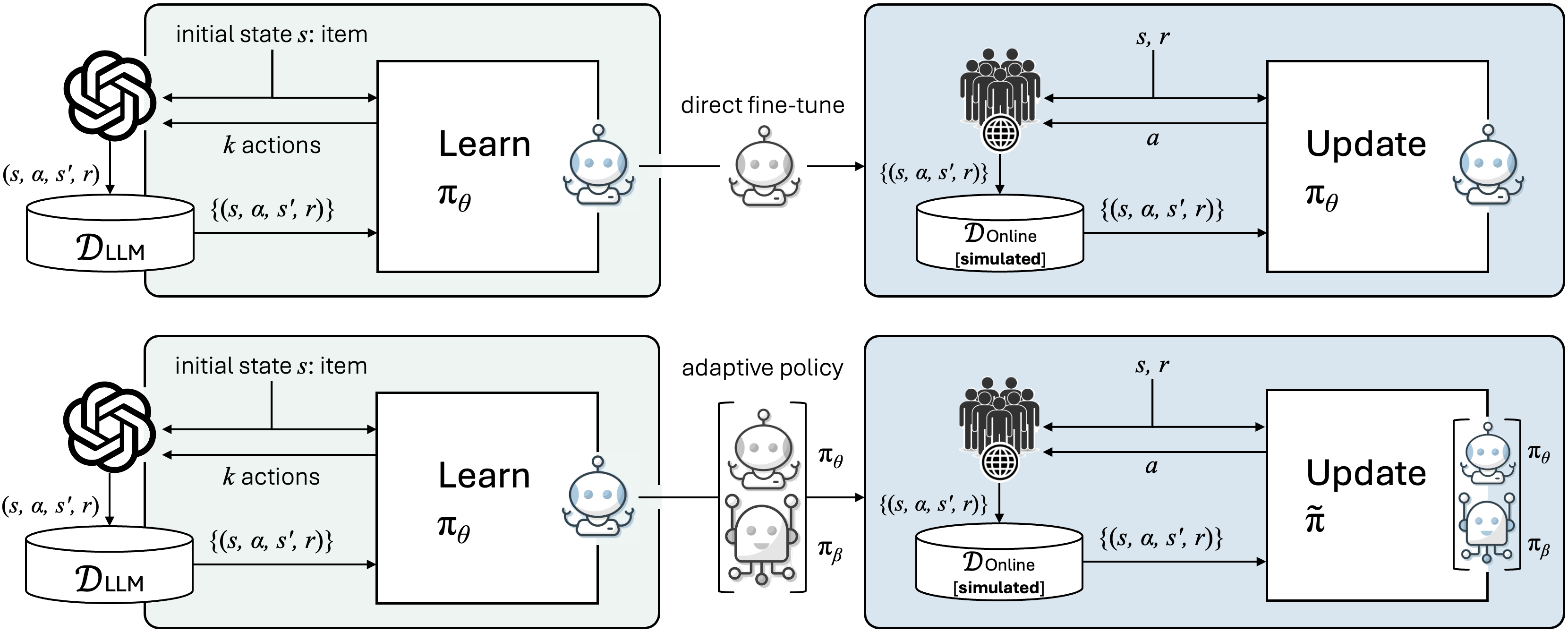}
        
        \label{figure_main_a}
        }
    \end{minipage}
    \hspace{0.2mm}
    \par
    \begin{minipage}[c]{0.78\textwidth}
    \centering
        \subfloat[A-iALP$_{ap}$: Adaptive policy combining $\theta$ and learnable policy $\beta$]{%
        \includegraphics[trim={0 0 0 340}, clip, scale=0.26]{Figures/Figures_WSDM25_Jie_05_new-2.png}
        \label{figure_main_b}
        }
    \end{minipage}
    \vspace{-0.5em}
    \caption{Illustration of different adaptation schemes of iALP-to-online. $D_{online}$ consists of generated actions $a$ from policy $\pi_{\theta}$ and corresponding rewards from the reward model. 
    }
    % }\vspace{-0.2cm}
    \label{figure_main}
\end{figure*}

\section{Method}
%In this section, 
We first introduce the interaction-Augmented Learned Policy (iALP), which is based on item preferences extracted from a Large Language Model. Following that, we outline strategies for adapting iALP (A-iALP) to an online RL-based recommendation system.

\subsection{Recommendation Policy trained on Preferences from LLM  }

\subsubsection{Preference Distillation}
\label{PD}
The principle of obtaining user preferences from an LLM involves using it as judge to evaluate which items %rom the candidate pool 
the user is likely to prefer or dislike, based on their interaction with the LLM. We design a preference prompt $p(s, a^k)$ to obtain this knowledge. 
 Table~\ref{prompt} presents the prompt template, which is guided by the description of the recommendation task for a specific scenario (e.g. scenario = track / video). The template uses the textual attributes of items (e.g., attribute1=title, attribute2=brand) to define the recommendation context. As illustrated in Figure~\ref{figure_preference}, the prompt contains the state $s$ and a list of $k$ (k=10 in the prompt example) actions $a^k$, which are fed to an LLM to provide a response, a choice $a$ from `a' to `k'. If the output is not None the corresponding action is selected for the user, otherwise the action is randomly selected from $a^k$.
We can see this procedure as an LLM-based sub-policy to select action:
\begin{gather}
\label{llm_action}
a \sim LLM[p(s, a^k)]
\end{gather}
Next, the reward $r^p$ for the action is designed as:
\begin{equation}\label{reward}
r^p = 
\begin{cases}
1.0 \ \ \text{ if the action is selected  } \\
0.0 \ \ \text{ if the response is None }  
\end{cases}
\end{equation}

\subsubsection{Policy Pre-training}
We utilize these responses as reward signals and training data to pre-train the recommendation policy offline. As the Figure~\ref{figure_preference} shows, once feedback from the LLM on actions is received, we train the interaction-Augmented Learned Policy (iALP) using the actor-critic (A2C) architecture \cite{konda1999actor}. This is combined with SASRec \cite{kang2018self} as a state encoder $G(\cdot)$, a widely adopted approach in RL-based RS. More specifically, given the interacted items $x_{1: t}$, the state at timestamp $t$ is:
\begin{gather}
s_t=G\left(x_{1: t}\right), 
\end{gather}
where $x_{1:t} = \{x_{1}, \ldots, x_{t}\}$ denote the interacted sequence of items, i.e., previous episode. $I$ denote the set of items in the system,  $x_{i} \in I(0<i \leq t)$ is the index of the interacted item ordered by timestamp.
$s_t$ is then used for the actor and critic networks:
\begin{gather}
\label{user_state}
a_t \sim \pi_{\theta}(\cdot|s_t),
\end{gather}
The implementation of $\pi_{\theta}$ contains two steps. First, $Actor$ network maps the state $s_t$ into action distribution, and a list of $k$ actions $a^k_t$ is sampled (random or top-$k$ actions), i.e., $a^k_t \sim Sample(Actor[s_t])$. Second, the preference distillation procedure in section~\ref{PD} is executed to select action $a_t$ and generate reward $r^p_t$, i.e. $a_t \sim LLM[p(s_t, a^k_t)]$.
The LLM can be seen as a helper policy to select the best action for the recommendation agent that finishes decision-making. 
Accordingly, the actor loss is formulated as:
\begin{equation}
\label{loss_L_hi}
\mathcal{L}_{A}= -\log \pi_{\theta}(a_t|s_t),
\end{equation}

Regarding the Q-learning network as critic network, we compute the Q-value as follows:
\begin{gather}
%\label{L_Q}
    Q(s_t, a_t) = Critic[s_t],
\end{gather}
The one-step time difference (TD) Q-learning loss as critic loss is defined as:
\begin{gather}
\label{L_Q}
\mathcal{L}_Q={(r^p(s_t, a_t)+\gamma \max _{a^{\prime}} Q(\mathrm{~s}_{t+1}, a^{\prime})-Q(\mathrm{~s}_t, a_t))^2},
\end{gather}
where $a_t$ is appended to $x_{1:t}$ to generate next state $s_{t+1}$ by $G(x_{1:t+1})$.
The advantage Q-value is calculated as:
\begin{gather}
    A(\mathrm{~s}_t, a_t)=r^p(s_t, a_t)+\gamma \max _{a^{\prime}} Q(\mathrm{~s}_{t+1}, a^{\prime})-Q(\mathrm{~s}_t, a_t),
\end{gather}
The actor loss $\mathcal{L}_A$ is finally formulated as  $\mathcal{L}_A \gets \mathcal{L}_A \cdot  A(\mathrm{~s}_t, a_t)$. In summary, the corresponding loss function for pre-training based on feedback from LLM is formulated as follows:
\begin{gather}
    \mathcal{L} = \mathcal{L}_A+\mathcal{L}_Q
\end{gather}

\subsection{Simulated Online Learning}
We propose two strategies to adapt iALP (A-iALP) to online recommendation for better long-term gains and alleviation of reward loss in the initial steps. Note that while in the previous phase we used the LLM as the reward and action model, here we do not utilise any LLM model. 

\subsubsection{Online Environment Simulation}
To provide immediate feedback and update online recommendation policies, we need to build RL environments. Due to the inaccessibility and high cost of real recommendation platforms, we use a simulated environment, following the approach of \citet{liu2023exploration} and \citet{shi2024large}. This simulated environment returns rewards based on state and action and can be easily constructed using public datasets. We employ two different simulated environments tailored to specific experimental settings, generating states based on user history and providing rewards for the recommended items (see Section~\ref{Datasets} for details).

We perform experiments on three user environments from different recommendation scenarios: \textbf{LFM}~\cite{schedl2016lfm}, \textbf{Industry}~\cite{ni2019justifying} and \textbf{Coat}\cite{schnabel2016recommendations}. The LFM is a music streaming platform Last.fm\footnote{http://www.last.fm/api} with textual content of title, album, and artist for each track (item). 
The Industry is the Industrial and Scientific category of Amazon review\footnote{https://nijianmo.github.io/amazon} with textual description of  title, category and brand for each product (item). 
We simulate an online environment for each scenario using a publicly available sequential interaction dataset, which is summarized in Table~\ref{dataset_statistics}.

\textbf{Reward Model}: For the Coat dataset, we follow the setting in \cite{shi2024large}, using Matrix Factorization (DeepFM)\cite{guo2017deepfm} to establish a user model, which acts as the reward model to simulate user feedback in training and testing environments. For the LFM and Industry datasets, we construct the reward model based on a sequential recommender method~\cite{liu2023exploration}, which generates scores/rewards based on a user's state consists of sequential items and the action. The Coat simulator focuses on user behaviors toward items, with the user as the initial state, while the latter two emphasizes the relationship between the recommended item and the user's interaction history, using an item as the initial state. 

\subsubsection{iALP-to-Online Fine-tuning}

Although online recommendation benefits from real user feedback for performance improvement, it is less sample efficient and experiences low returns at the beginning of training.
In contrast, the pre-training RL policy is sample-efficient since no online interactions are required. 
Therefore, instead of treating online recommendation as an individual topic, we directly use the policy $\pi_\theta$ learned from LLM to generate the online policy and, in particular, we sample the action from policy $\pi_\theta$:
\begin{gather}
   a_t \sim \pi_{\theta}(\cdot|s_t) = Actor[s_t], 
\end{gather}
The use of the pre-trained policy $\pi_\theta$ essentialy means that we deploy the pre-trained actor-critic network with the help of the simulator that provides the corresponding rewards $r$.
The policy is then is fine-tuned based on the simulated real-time user reward $r$ (feedback), we call this method A-iALP$_{ft}$.
The online actor loss and critic loss are as follows:
\begin{gather}
\mathcal{L}^{on}_Q={(r(s_t, a_t)+\gamma \max _{a^{\prime}}
Q(\mathrm{~s}_{t+1}, a^{\prime})-Q(\mathrm{~s}_t, a_t))^2}, \\
\mathcal{L}^{on}_{A}= -\log \pi_{\theta}(a_t|s_t),
\end{gather}
The key difference of A-iALP$_{ft}$ is that the same algorithm is further improved for online learning using the simulated environment reward $r$ to replace the designed signal $r_p$ in Eq.~\ref{L_Q}. 
The process is also illustrated in Figure~\ref{figure_main_a}, which directly uses online feedback to continue to optimize  $\pi_\theta$.

\begin{algorithm}[h]
    \caption{Training iALP Phase}
    \label{alg: llm rl}
    \begin{algorithmic}[1]
        \STATE {\bf Require}: Item set $I$  (action space $\mathcal{A}$), LLM
        \STATE Initialize replay buffer $\mathcal{D}_{\rm LLM}$ by initial user sequence $x_1\in I$
        \STATE Initialize parameters of the agent $\theta$: actor network $Actor$, critic network $Critic$, and the state encoder $G$
        \FOR{each iteration}
        \STATE Obtain initial user sequence $x_1$ from $\mathcal{D}_{\rm LLM}$
        \FOR{step $t=1, \cdots, T$}
        \STATE Generate state $s_t$ by $G(x_{1:t})$
        \STATE Execute $Actor[s_t]$ to produce action distribution
        \STATE Sample a list of $k$ actions $a^k_t$
        \STATE Construct prompt $p(s_t, a^k_t)$ based on template
        \STATE Act $LLM(p(s_t, a^k_t))$
        to generate $a_t (i.e., x_{t+1}), r_t$ %to produce $s_{t+1}$
        \STATE Store transition $(x_{1:t}, a_t, r_t, x_{1:t+1})$ in $\mathcal{D}_{\rm LLM}$
        \STATE Sample minibatch $b$ from $\mathcal{D}_{\rm LLM}$
        \STATE With $b$, calculate $\mathcal{L}_Q$ and $\mathcal{L}_A$
        \STATE Update $\theta$
        \ENDFOR
        \ENDFOR 
        \STATE \textbf{Return} Agent $\theta$: $G$, $Actor$, and $Critic$
    \end{algorithmic}
\end{algorithm}

\subsubsection{Adaptive Policy}

Although direct fine-tuning is straightforward, it presents several challenges. For example, the pre-trained policy might be compromised or even deteriorate during the initial phase of online training, particularly due to distribution shifts between LLM-generated preferences and actual online user behavior~\citep{balanced_replay}. Another concern is that exploration~\citep{anti_exploration} may be overlooked as the iALP often prioritizes non-preferred actions, limiting the system's ability to discover newer and potentially better recommendations.

To solve the above issues, we propose an alternative scheme A-iALP$_{ap}$ (illustrated in Figure~\ref{figure_main_b}) that combines the existing algorithm with the online policy to allow further learning. Given a policy $\pi_{\theta}$ obtained from pre-training phase, instead of
directly fine-tuning the parameters,
we freeze $\pi_{\theta}$ and propose another learnable policy $\pi_{\beta}$. Both are added into a policy set % $\Pi$:
$\tilde\pi=[\pi_{\theta}, \pi_{\beta}]$,
which is responsible for further performance improvement during online training.
The final policy $\tilde{\pi}$ can be represented  as follows:
\begin{gather}\label{eq:rex_policy}
    a_t \sim \tilde{\pi}(\cdot|s_t) =  (1-\alpha)\pi_{\theta}(\cdot|s_t) + \alpha\pi_{\beta}(\cdot|s_t), \\
    \mathcal{L}^{on}_{A}= -\log \tilde{\pi}(a_t|s_t)
\end{gather}
where $\alpha$ is the weight of taking actions from the learnable policy $\pi_{\beta}$, $1-\alpha$ indicates the probability of using the pre-trained policy $\pi_{\theta}$. As the training steps increase, $\alpha$ increases to 1. In practice, the initial value of $\alpha$ depends on the specific recommendation scenario.
It is intuitive to understand that the pre-trained policy $\pi_{\theta}$ mainly determines actions in the early stages of training, and as the new policy $\pi_{\beta}$ learns more and more preferences, it will take over the decision-making process at the end of the fine-tuning process. 

\subsection{Discussion}
Algorithms~\ref{alg: llm rl} and ~\ref{alg: on rl} illustrate the iALP pre-training and iALP-to-Online procedures, respectively.
Note that the critic and state encoder networks of A-iALP$_{ap}$ are updated the same way as A-iALP$_{ft}$. The total loss of online recommendations is $\mathcal{L}^{on}=\mathcal{L}^{on}_{A}+\mathcal{L}^{on}_{Q}$. We can retain the useful behaviours of iALP to generate desirable sequences in the initial online training stage.
The adaptive scheme enables learning new abilities by interact with the user environment.

\begin{algorithm}[h]
    \caption{A-iALP: iALP-to-Online Phase}
    \label{alg: on rl}
    \begin{algorithmic}[1]
        \STATE {\bf Require}:  iALP $\theta$, learnable agent $\beta$, scheme, weight $\alpha$
        \STATE Initialize empty online replay buffer $\mathcal{D}_{\rm Online}$
        \STATE Initialize parameters of the agent $\beta$ with $\theta$.
        \FOR{each iteration}
        \STATE Obtain initial state $s_0$ from environment
        \FOR{step $t=1, \cdots, T$}
        \IF{scheme is A-iALP$_{ft}$}
        \STATE Act $a_t \sim \pi_\theta(\cdot|s_t)$ %to produce action $a_t$
        \ELSIF{scheme is A-iALP$_{ap}$}
        \STATE Act $a_t \sim (1-\alpha)\pi_\theta(\cdot|s_t) + \alpha \pi_\beta(\cdot|s_t)]$ 
        \ENDIF
        \STATE Obtain $r_t$, $s_{t+1}$ from environment
        \STATE Store transition $(s_t, a_t, r_t, s_{t+1})$ in $\mathcal{D}_{\rm Online}$
        \STATE Sample minibatch $b$ from $\mathcal{D}_{\rm Online}$
        \STATE With $b$, calculate $\mathcal{L}^{on}_Q$ and $\mathcal{L}^{on}_A$
        \STATE Update $\beta$
        \ENDFOR
        \ENDFOR \\
    \end{algorithmic}
\end{algorithm}

\section{Experimental Setup}
\subsection{Research Questions} 
\label{RQ}
In this section, we describe the experimental setup used to validate our proposed A-iALP approach.
We aim to answer the following research questions:
\\
\textbf{RQ1:} How does iALP compare to traditional purely online RL-based recommendation methods in the initial stages of online recommendation?
\\
\textbf{RQ2:} How effective is A-iALP at achieving long-term effectiveness in online recommendations?
\\
\textbf{RQ3:} 
How does A-iALP compare to directly incorporating preference of LLMs in online training?
\\
\textbf{RQ4:}  How does A-iALP perform under various exploration-exploitation strategy?

Table 2 summarises the three data sets (section 4.2.1) used for the experiments. The datasets are partitioned to simulate online training (80\% dataset) and testing (20\% dataset) environments.

\label{Datasets}

\begin{table}[] %\footnotesize
%\fontsize{9}{10}\selectfont
\setlength{\tabcolsep}{2.5pt}
\caption{Dataset statistics. 
}
\label{dataset_statistics}
\centering
\begin{tabular}{lcccl}

\toprule
Dataset    & \#item          & \#seq./user  &\#inter.   &Item attribute       \\ 
\midrule
LFM      & 18,297         &11,073     &146,255 &\small title; album; artist \\
Industry   &5,814         &10,935   &71,872 & \small title; category; brand \\
Coat    &290        &300   &11,600 & \small jacket colour; type \\
\bottomrule
\end{tabular}
\vspace{-0.4cm}
\end{table}

\subsection{Baselines}
We compare the proposed A-iALP with several online RL methods for online recommendation.
\begin{itemize} [leftmargin=*]
\item DQN~\cite{mnih2013playing}: A Q-learning based off-policy method learning from data produced by a policy different from the one currently being optimized.
    \item PG~\cite{williams1992simple}: An on-policy gradient-based method that utilizes data generated
by their current policy. 
    \item A2C~\cite{konda1999actor}: An on-policy gradient-based method with the actor-critic framework, which is the same as our proposed iALP in the pre-training stage.
    \item iALP: The pre-trained agent to take advantage of the preferences derived from LLM and optimised based on A2C. This model serves as a baseline, providing recommendations without further exploration or updates.
\end{itemize}
The two adaptation strategies of our A-iALP are described as A-iALP$_{ft}$ and A-iALP$_{ap}$, respectively, for clarity.
\subsection{Metrics} 
We apply three metrics: Return, Length and Average Reward\cite{yu2024easyrl4rec,seo2021state} to evaluate the RL-based RSs.
Return ($R=\sum_t r_t$ ) measures the cumulative rewards (return)
of the recommended episode/sequence, which is applied in the RL methods~\cite{seo2021state} to evaluate long-term gains. The higher Return denotes better performance.  In addition, the length ($Len$) of the sequence and the mean reward ($R_{avg}=R/Len$) of the actions are also used as a reference~\cite{yu2024easyrl4rec,shi2024large} to analyse user engagement and performance of the immediate recommendation of the policy. 
To evaluate in early training steps, we incorporate $R@e$, $Len@e$ and $R_{avg}@e$, where $e \in \{0, 1, 2\}$, to show the results in epoch $e$.

\subsection{Implementation Details} 
For the training stage of iALP, to generate the correct label format, we first tune the LLM with the commonly used LORA in the PEFT~\cite{hu2021lora} strategy to learn the template prompt via 1000 randomly sampled interactions. 
The LLM we use is Mistral 7B~\cite{jiang2023mistral}.
iALP is an A2C framework trained with generated preferences (action, reward) from LLM at every step for 100 epochs.
In the online stage, we use the same architecture for A-iALP. The difference in A-iALP$_{ap}$ is that it freezes the iALP policy network and initialises randomly another learnable policy with the same network. For online learning, all RL methods are trained with 50k steps and evaluated in the same test environment, and the most recent 20 episodes are used to generate the learning curve results. 

\section{Experimental Results}
In this section, we present the experimental results to answer the research questions described in Section~\ref{RQ}.
\subsection{Initial Performance Comparison (RQ1)}
Table~\ref{r_0} shows the performance of directly applying iALP to recommend items without learning from online environment, i.e., e=0.  
We see that iALP significantly outperforms traditional on-line RL methods that are randomly initialised in the context of cold-start online recommendation. The results demonstrate that iALP, which is trained on preferences distilled from LLMs, can provide a relatively desirable item sequence in terms of cumulative rewards and sequence length of the recommended items. Specifically, iALP manages to alleviate the common issue of poor quality recommendations during the initial phase of deployment, which is a typical problem for traditional RL methods trained from scratch. This advantage not only improves the user experience from the outset, but also accelerates the process of online updates and learning. By leveraging the pre-trained knowledge, iALP is able to bypass the initial random exploration phase that purely online RL methods undergo, which often results in suboptimal recommendations and user dissatisfaction. Instead, iALP starts with a more informed and refined strategy, leading to higher initial user engagement and satisfaction. %We can also see from Table~\ref{r_1} and ~\ref{r_2}, 
We also see in Table \ref{r_1}% and \ref{r_2} 
that, after incorporating a small amount of user feedback (training 1 
%or 2 
epoch), A-iALP maintains greater long-term returns compared to general methods. This demonstrates that A-iALP leverages effectively user feedback to enhance performance. This improved starting point facilitates gathering more relevant early feedback, enabling faster and more effective online learning and adaptation.
\begin{table}[h!]
%\fontsize{8}{9}\selectfont
%\setlength{\tabcolsep}{4.8pt}
\centering
\caption{The initial performance, i.e., epoch=0, on LFM and Industry environments for online RL-based recommendation. Boldface denotes the highest score.}
\label{r_0}
%\vspace*{-4mm}
\begin{tabular}{p{1.2cm}<{\centering}p{0.7cm}<{\centering}p{0.7cm}<{\centering}p{0.7cm}<{\centering}p{0.002cm}p{0.7cm}<{\centering}p{0.7cm}<{\centering}p{0.7cm}<{\centering}} 
\toprule
\multirow{2}{*}{Method} & \multicolumn{3}{c}{LFM} & & \multicolumn{3}{c}{Industry} \\
\cmidrule{2-4} \cmidrule{6-8}
& $R@0$ & $Len@0$ & $R_{avg}@0$ && $R@0$ & $Len@0$ & $R_{avg}@0$ \\
\midrule
DQN &1.53&4.52&0.34 & &1.12&5.21 &0.21 \\
PG &1.73&4.73&0.37 & &1.06 &5.15 &0.22\\
A2C &1.96&5.15&0.38 & &2.43 &5.92 &0.41\\
iALP &\textbf{5.11}&\textbf{6.21}&\textbf{0.82} & &\textbf{6.35} &\textbf{5.94} &\textbf{1.06} \\
\bottomrule
\end{tabular}
\end{table}

\begin{table}[h!]
%\fontsize{8}{9}\selectfont
%\setlength{\tabcolsep}{4.8pt}
\centering
\caption{Performance of online RL-based recommendation on LFM and Industry environments while epoch=1. Boldface denotes the highest score.}
\label{r_1}
%\vspace*{-4mm}
\begin{tabular}{p{1.3cm}<{\centering}p{0.65cm}<{\centering}p{0.7cm}<{\centering}p{0.7cm}<{\centering}p{0.002cm}p{0.65cm}<{\centering}p{0.7cm}<{\centering}p{0.7cm}<{\centering}} 
\toprule
\multirow{2}{*}{Method} & \multicolumn{3}{c}{LFM}  && \multicolumn{3}{c}{Industry} \\
\cmidrule{2-4} \cmidrule{6-8}
& $R@1$ & $Len@1$ & $R_{avg}@1$ && $R@1$ & $Len@1$ & $R_{avg}@1$ \\
\midrule
DQN &5.04&6.01&0.84&&3.84&6.51&0.59 \\
PG &4.85&6.32&0.77&&3.32&5.96&0.56\\
A2C &6.12&6.84&0.89&&7.32&6.41&\textbf{1.14}\\
A-iALP$_{ft}$ &\textbf{8.83}&\textbf{8.01}&\textbf{1.11} &&\textbf{9.28}&\textbf{8.62}&1.07 \\
\bottomrule
\end{tabular}
\end{table}

% \begin{table}[h!]
% %\fontsize{8}{9}\selectfont
% %\setlength{\tabcolsep}{4.8pt}
% \centering
% \caption{Performance of online RL-based recommendation on LFM and Industry environments while epoch=2. Boldface denotes the highest score.}
% \label{r_2}
% %\vspace*{-4mm}
% \begin{tabular}{p{1.3cm}<{\centering}p{0.65cm}<{\centering}p{0.7cm}<{\centering}p{0.7cm}<{\centering}p{0.002cm}p{0.65cm}<{\centering}p{0.7cm}<{\centering}p{0.7cm}<{\centering}} 
% \toprule
% \multirow{2}{*}{Method} & \multicolumn{3}{c}{LFM} & & \multicolumn{3}{c}{Industry} \\
% \cmidrule{2-4} \cmidrule{6-8}
% & $R@2$ & $Len@2$ &$R_{avg}@2$ && $R@2$ & $Len@2$ & $R_{avg}@2$ \\
% \midrule
% DQN &8.32&7.43&1.12&&16.3&8.32&1.96 \\
% PG &8.23&7.23&1.14&&14.5&8.43&1.72\\
% A2C &8.39&8.04&1.04&&17.2&8.36&2.06\\
% A-iALP$_{ft}$ &\textbf{14.13}&\textbf{8.52}&\textbf{1.66} &&\textbf{22.1}&\textbf{9.43}&\textbf{2.34} \\
% \bottomrule
% \end{tabular}
% \end{table}

\subsection{Long-term Performance Comparison (RQ2)}

\begin{figure}[!t]
    \captionsetup[subfloat]%{captionskip=-5pt,nearskip=0pt,farskip=0pt}
    {}
    \centering
    \begin{minipage}[t]{0.23\textwidth}
    \centering
        %\subfloat[]{%
        \includegraphics[trim={5 0 5 0}, clip, scale=0.39]{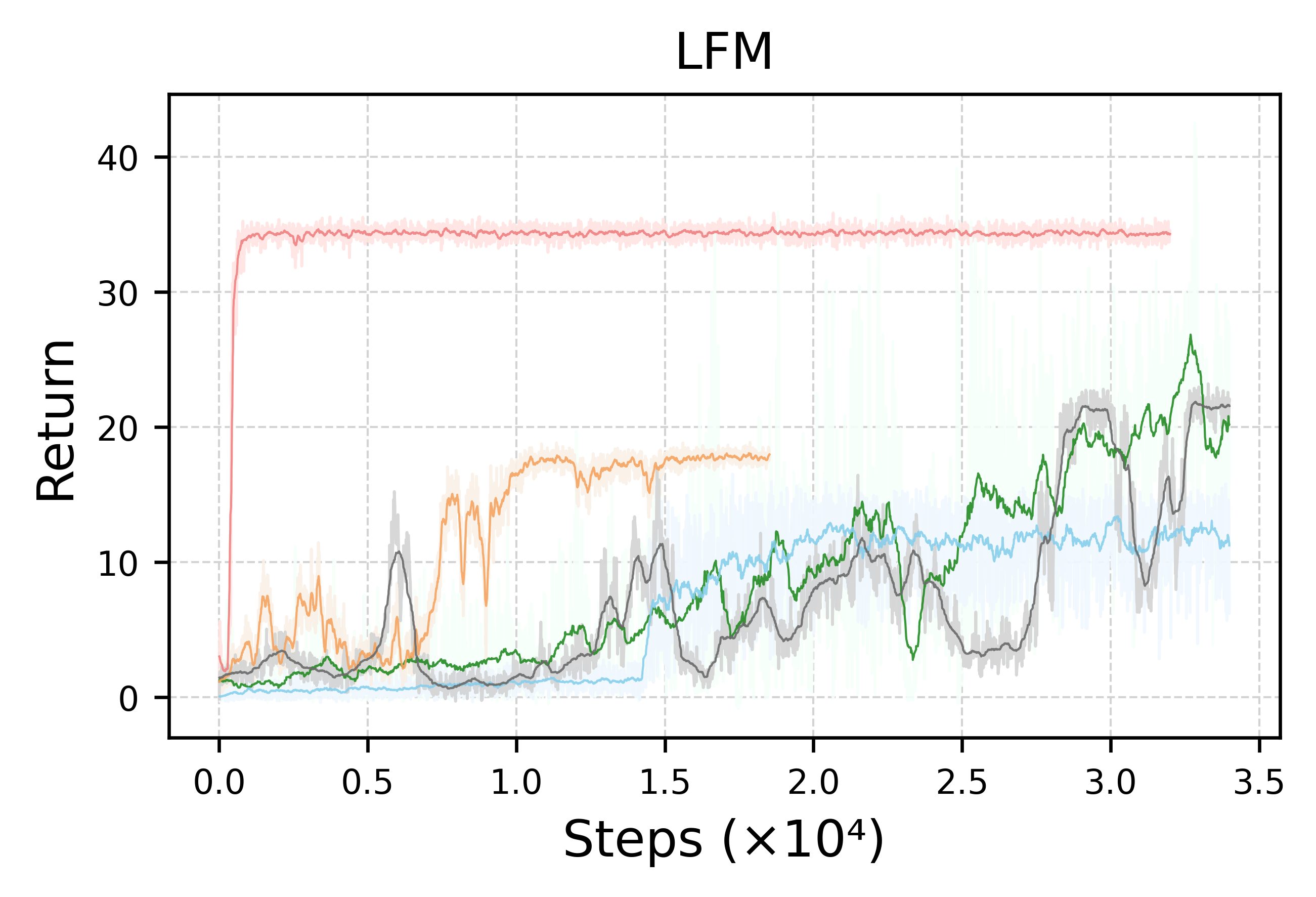}
        \label{figure1_a}
        %}
        \phantomcaption
    \end{minipage}
    %\hspace{0.2mm}
    \hspace{0.07mm}
    \begin{minipage}[t]{0.23\textwidth}
    \centering
        %\subfloat[]{%
        \includegraphics[trim={5 0 5 0}, clip, scale=0.39]{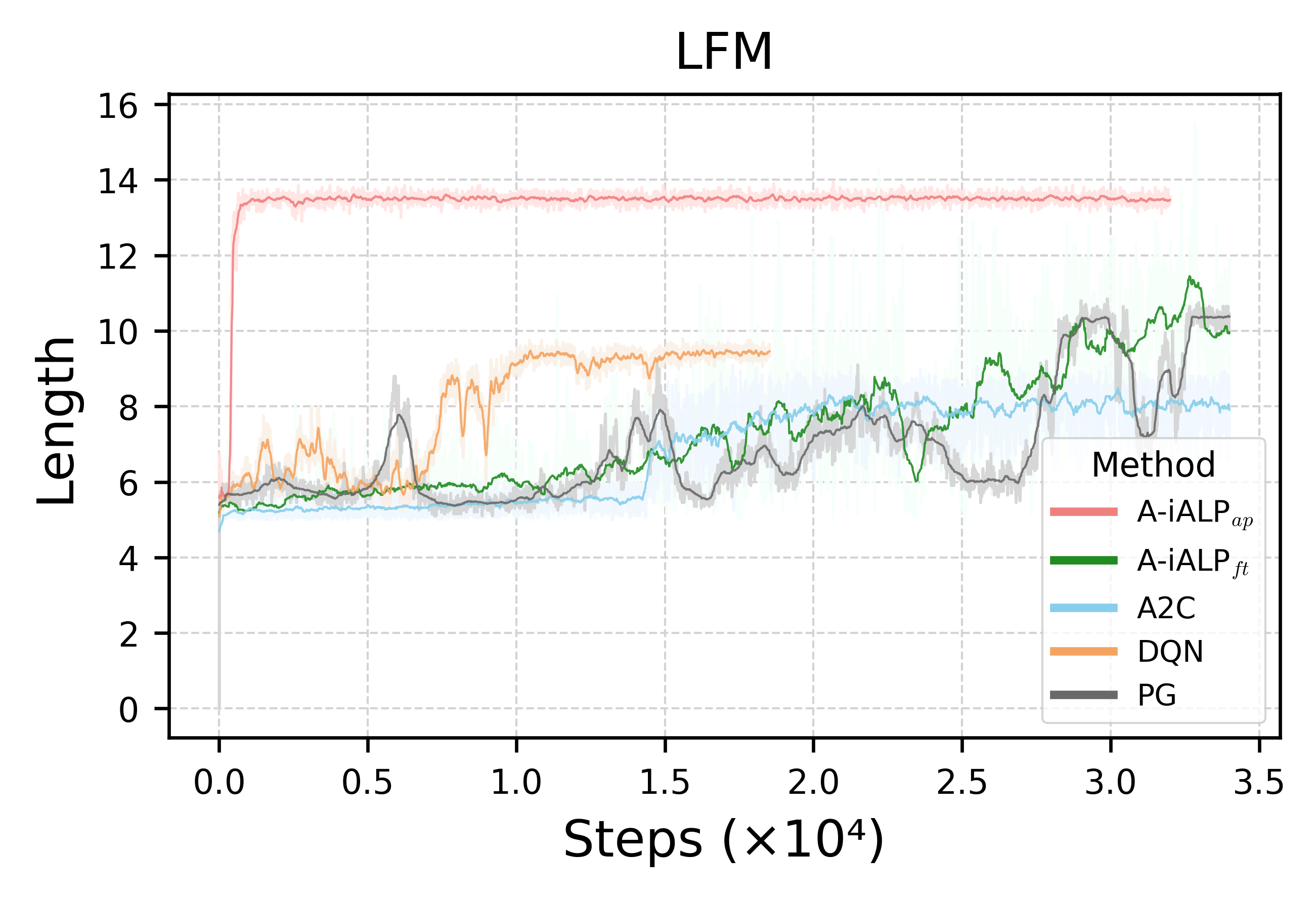}
        \label{figure1_b}
        %}
        \phantomcaption
    \end{minipage}
    \vspace{-0.5cm}
    \caption{Learning curves of return (left) and length (right) between baselines and A-iALP on LFM.}
    % \vspace{-0.5cm}
    \label{figure_RQ2_lfm}
\end{figure}

\begin{figure}[!t]
    \captionsetup[subfloat]%{captionskip=-5pt,nearskip=0pt,farskip=0pt}
    {}
    \centering   
    \begin{minipage}[t]{0.23\textwidth}
    \centering
        %\subfloat[]{%
        \includegraphics[trim={5 0 5 0}, clip, scale=0.39]{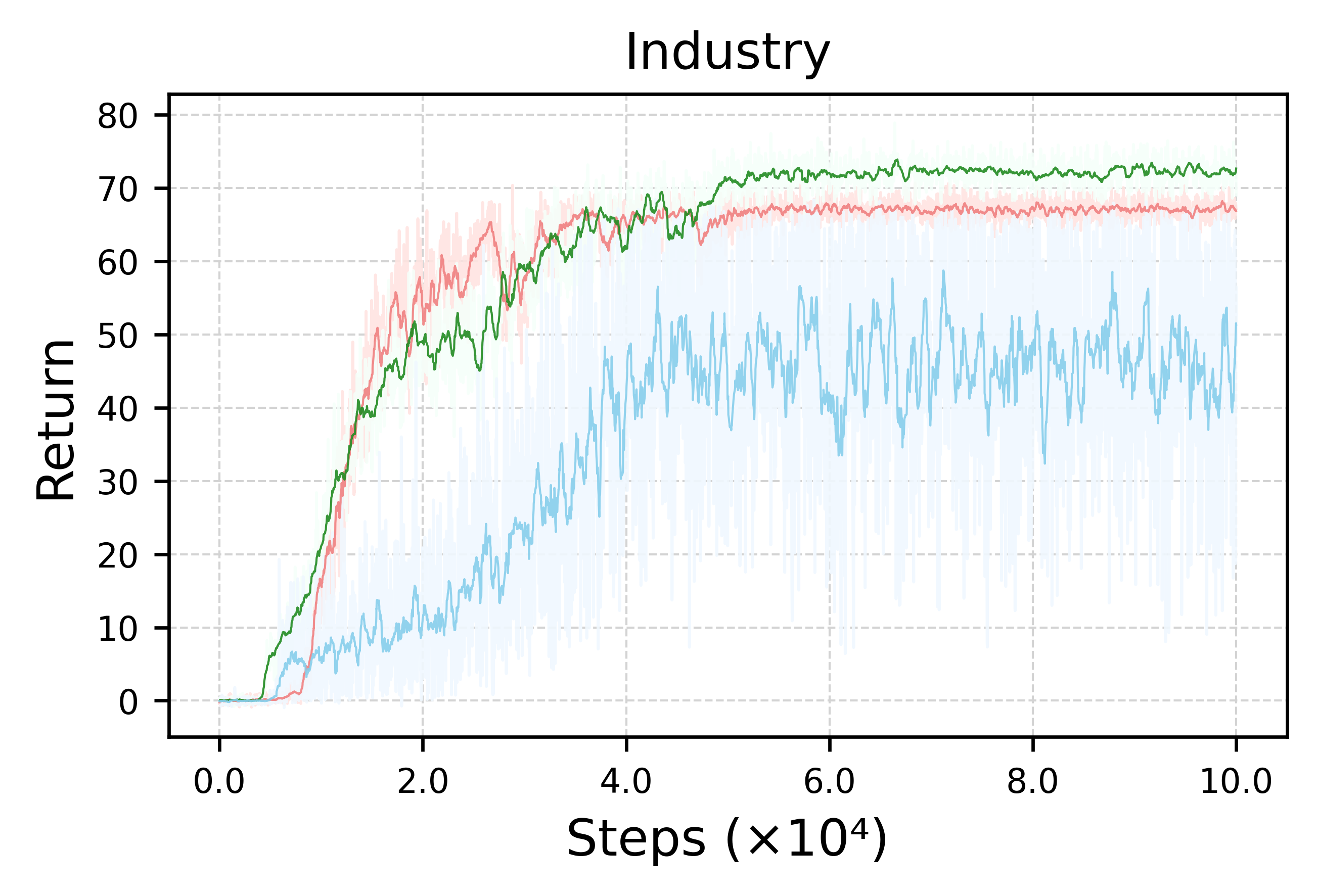}
        \label{figure1_b}
        %}
    \end{minipage}
    \hspace{0.07mm}
    \begin{minipage}[t]{0.23\textwidth}
    \centering
        %\subfloat[]{%
        \includegraphics[trim={5 0 5 0}, clip, scale=0.39]{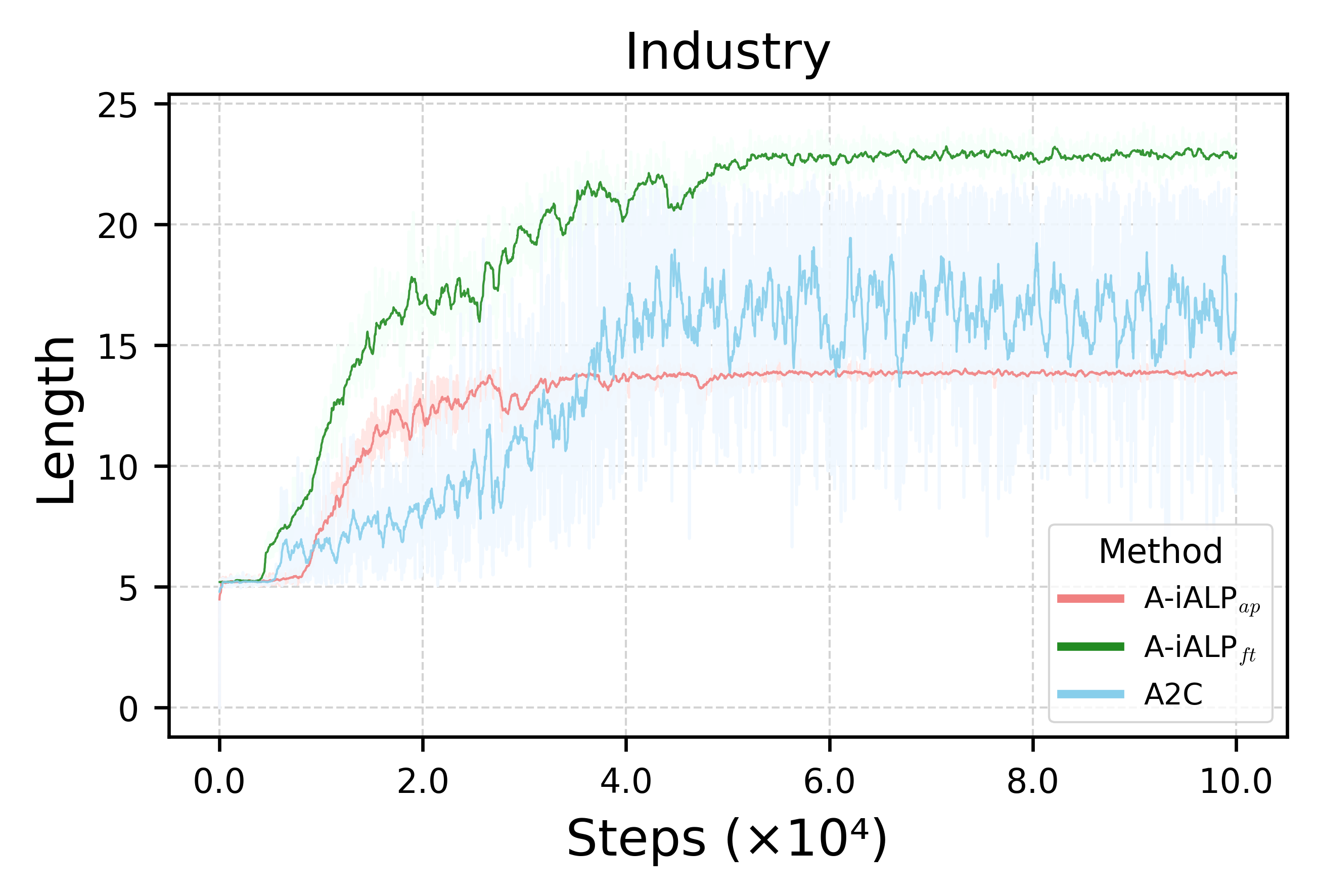}
        \label{figure1_b}
        %}
    \end{minipage}
    \vspace{-0.5cm}
    \caption{Learning curves of return (left) and length (right) between several baselines and A-iALP on Industry.}
    % }\vspace{-0.2cm}
    \label{figure_RQ2_industry}
\end{figure}

Figures~\ref{figure_RQ2_lfm} and Figure~\ref{figure_RQ2_industry} illustrate the training curves between baseline and A-iALP w.r.t. returns, the sequence length on the LFM and Industry, respectively.
The performance of all methods are presented on the LFM, while Industry focusses on the top three methods.
We can see that in both LFM and Industry A-iALP$_{ap}$ converges quickly and demonstrates stability, which achieves best return at around 1,000 steps and 20,000 step, respectively. In Industry, A-iALP$_{ft}$ performs second best on Industry before 40,000 environmental steps, while A-iALP$_{ap}$ takes the first position as steps continue. In LFM, A-iALP$_{ft}$ increases stably though with slightly lower performance at the beginning.
The performance in the test environment from is illustrated in Table~\ref{RQ2}.
We can observe that A-iALP in general outperforms the training from scratch methods (DQN, PG and A2C) in return and length. In the metric of length, there is no obvious differences on all methods, which might be attributed to the phenomenon that the length of recommended sequence might not reflect the user's long-term satisfaction. Specifically, A-iALP$_{ft}$ shows large improvements over iALP, implying the benefits brought about by additional online training over pure offline training based on LLM interactions. The proposed A-iALP$_{ap}$ outperforms baselines method overall. It shows stable improvements starting from a desirable performance brought about by the pre-training procedure. 

\begin{table}[h!]
%\fontsize{8}{9}\selectfont
%\setlength{\tabcolsep}{4pt}
\centering
\caption{Performance comparison of online RL methods for online recommendation. Boldface denotes the highest score, and the second-best results are underlined.}
\label{RQ2}
%\vspace*{-4mm}
\begin{tabular}{p{0.45cm}p{0.45cm}<{\centering}p{0.65cm}<{\centering}p{0.65cm}<{\centering}p{0.65cm}<{\centering}p{0.65cm}<{\centering}@{\hspace{3pt}}p{1.2cm}<{\centering}@{\hspace{4pt}}p{1.2cm}<{\centering}} 
\toprule
&&\small DQN &\small PG & \small A2C &\small iALP &\small A-iALP$_{ft}$ &\small A-iALP$_{ap}$ \\
\midrule
\multirow{3}{*}{LFM}&$R$ &28.8 &29.7 &28.1 &11.2 &31.5 &\textbf{33.1} \\
&$Len$ &9.53 &8.46 &9.29 &7.21 &10.1 &\textbf{10.2}\\
&$R_{avg}$ &3.02 &3.51 &1.55 &3.21 &3.13 &\textbf{3.22}\\
\bottomrule
\multirow{3}{*}{{\shortstack{Indu-\\stry}}}&$R$ &40.3 &43.3 &46.3 &25.3 &\textbf{52.4} &51.8 \\
&$Len$ &11.3 &10.4 &10.8 &10.7 &\textbf{11.5} &11.3\\
&$R_{avg}$ &3.57 &4.16 &4.28 &2.36 &4.55 &\textbf{4.58}\\
\bottomrule
\multirow{3}{*}{Coat}&$R$ &54.3 & 79.3 &81.7 &31.2&83.2&\textbf{84.4} \\
&$Len$ &22.1&29.8& \textbf{29.9}&12.9& 29.6& 29.7\\
&$R_{avg}$ &2.47 &2.66 &2.73&2.42&2.81&\textbf{2.84}\\ 
\bottomrule
\end{tabular}
\end{table}

\subsection{Utilisation of LLM (RQ3)}

\begin{figure}[!t]
    \captionsetup[subfloat]%{captionskip=-5pt,nearskip=0pt,farskip=0pt}
    {}
    \centering
    \begin{minipage}[c]{0.23\textwidth}
    \centering
        %\subfloat[]{%
        \includegraphics[trim={0 20 0 0}, clip, scale=0.37]{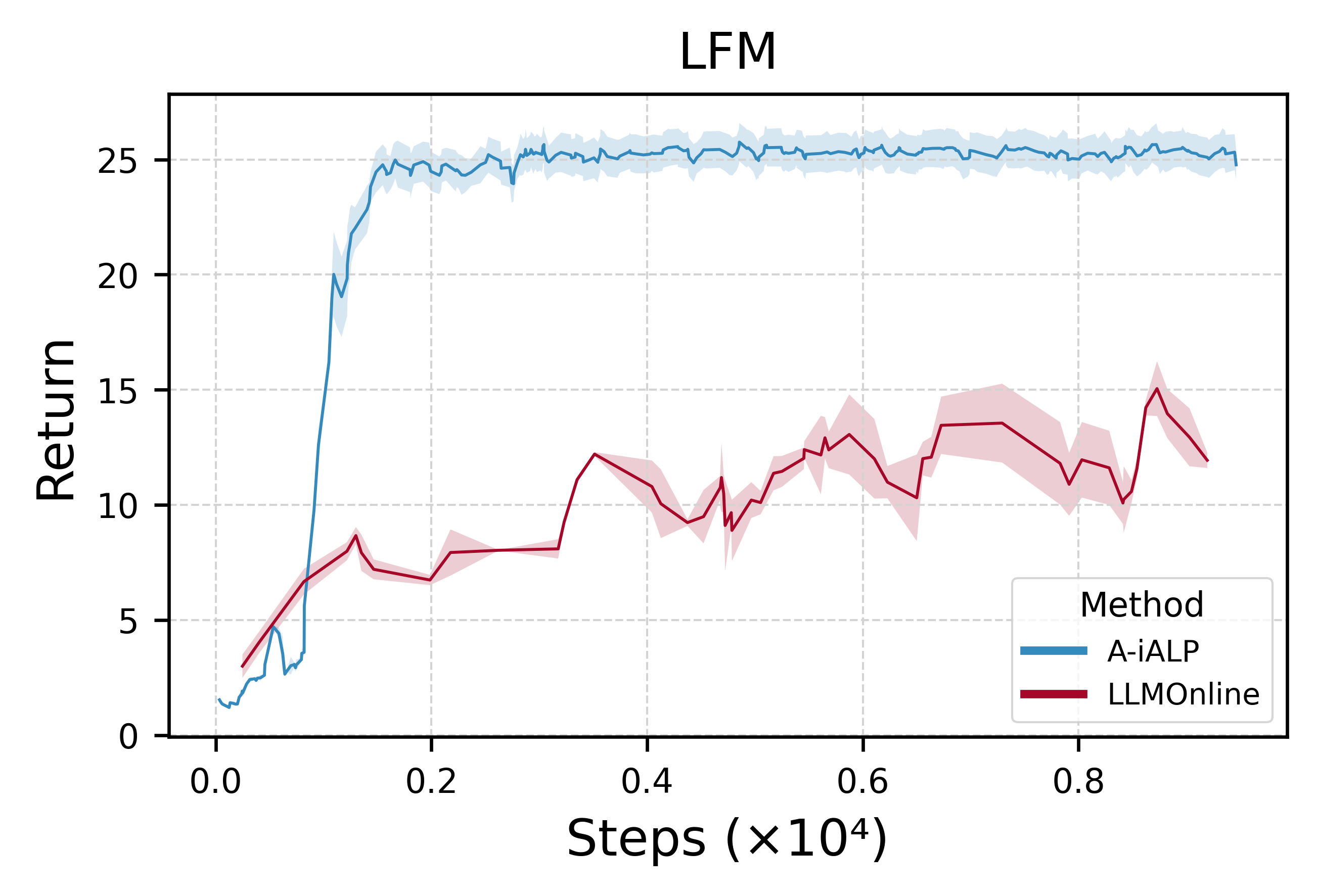}
        \label{figure1_a}
        %}
    \end{minipage}
    %\hspace{0.2mm}
    \hspace{0.07mm}
    \begin{minipage}[c]{0.23\textwidth}
    \centering
        %\subfloat[]{%
        \includegraphics[trim={0 20 0 0}, clip, scale=0.37]{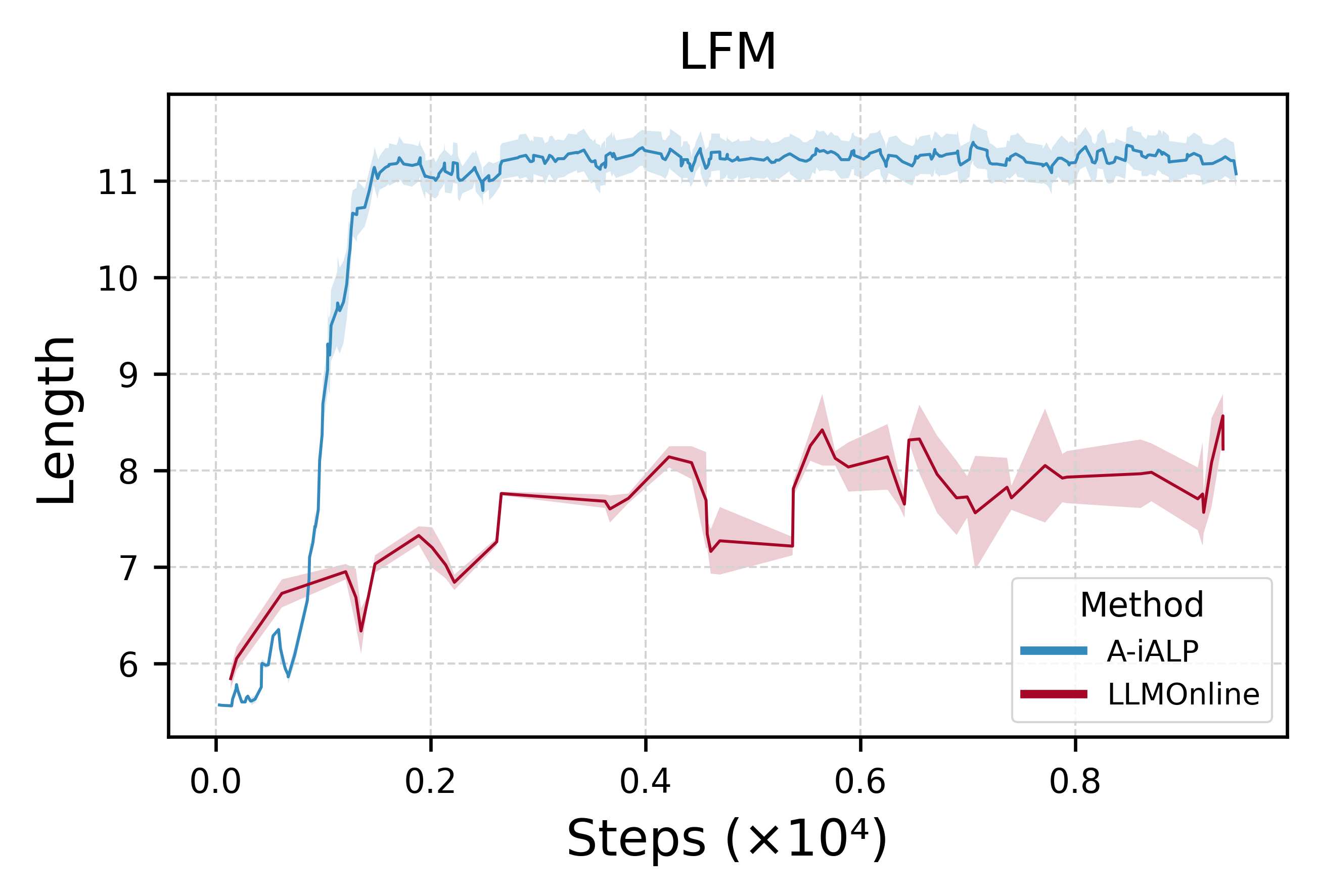}
        \label{figure1_b}
        %}
    \end{minipage}
    \vspace{-0.5cm}
    \caption{Comparison of return (left) and length (right)
between using LLM online and A-iALP method on LFM.}
    % }\vspace{-0.2cm}
    \label{figure_RQ3_lfm}
\end{figure}

\begin{figure}[!t]
    \captionsetup[subfloat]%{captionskip=-5pt,nearskip=0pt,farskip=0pt}
    {}
    \centering
    \begin{minipage}[c]{0.23\textwidth}
    \centering
        %\subfloat[]{%
        \includegraphics[trim={0 20 0 0}, clip, scale=0.37]{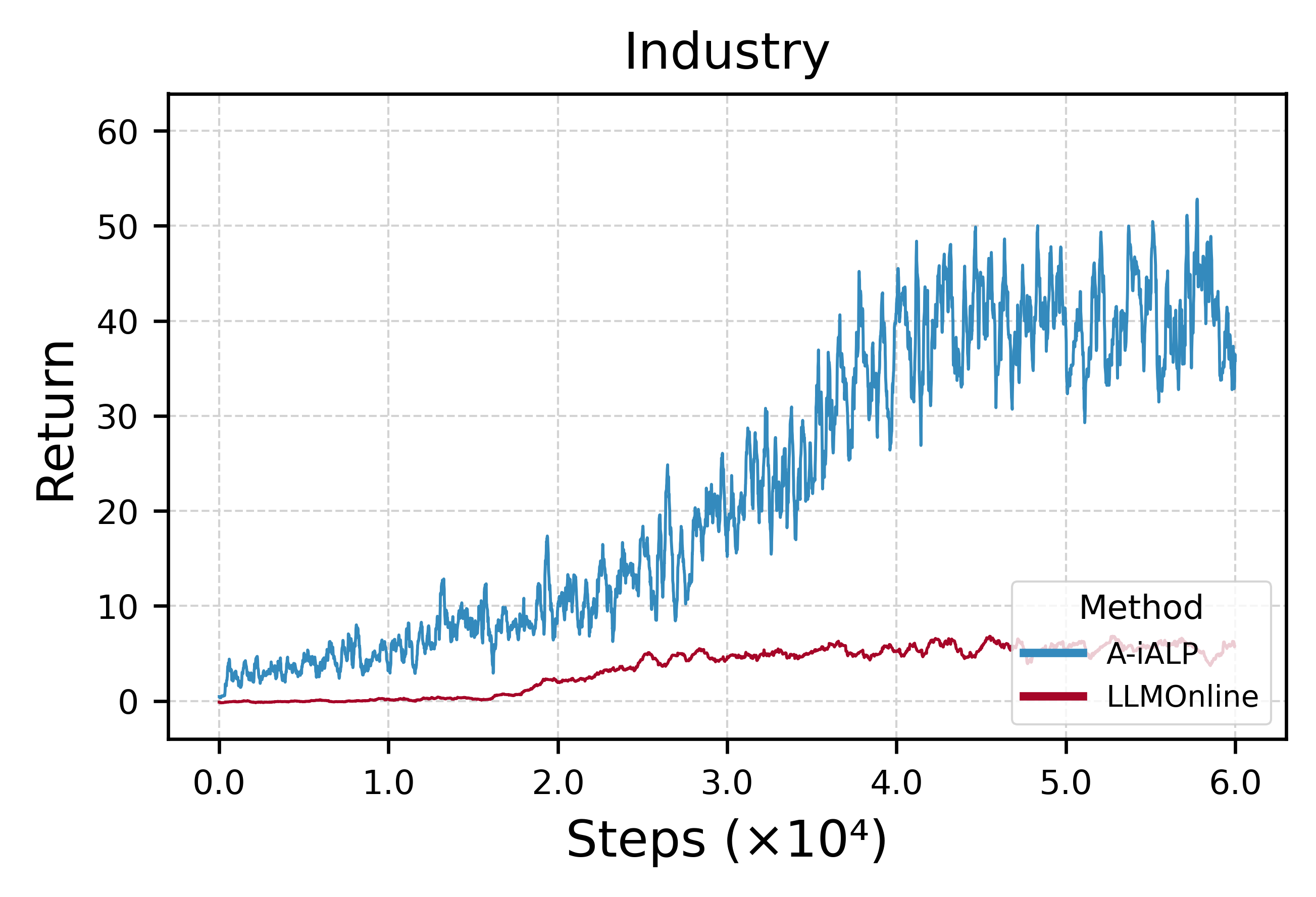}
        \label{figure1_a}
        %}
    \end{minipage}
    %\hspace{0.2mm}
    \hspace{0.07mm}
    \begin{minipage}[c]{0.23\textwidth}
    \centering
        %\subfloat[]{%
        \includegraphics[trim={0 20 0 0}, clip, scale=0.37]{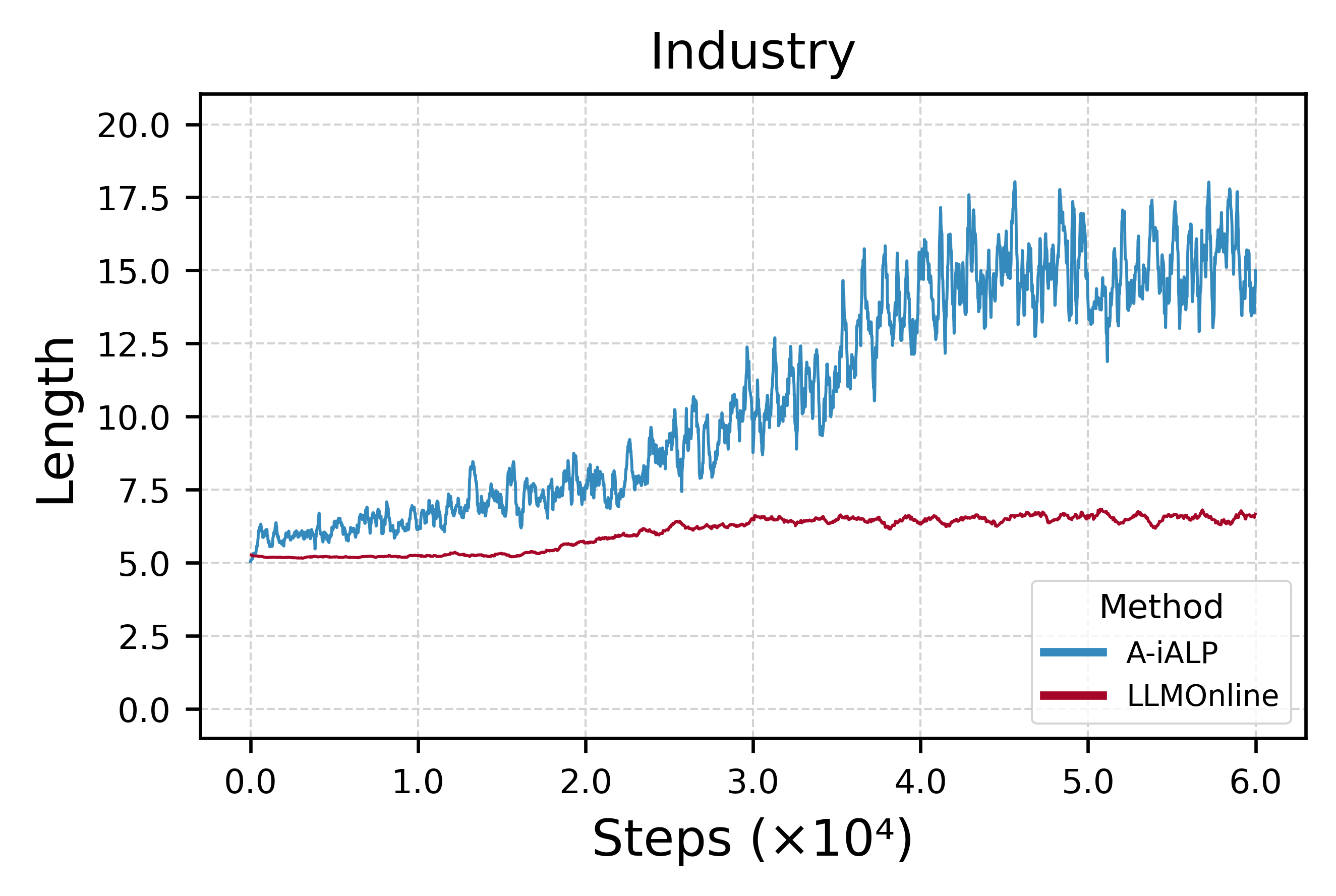}
        \label{figure1_b}
        %}
    \end{minipage}
    \vspace{-0.5cm}
    \caption{Comparison of return (left) and length (right)
between using LLM online and A-iALP method on Industry.}
    % \vspace{-0.2cm}
    \label{figure_RQ3_industry}
\end{figure}

Here we compare A-iALP with LLMOnline, which directly uses LLM as a sub-agent in an online environment for taking actions.
The goal is to verify the effectiveness of our iALP-to-Online recommendation procedure. LLMOnline is similar to pre-training iALP, but differs in that the agent is directly updated by user feedback in the online RL-based recommendation, rather than by a predefined reward function. 
The results on LFM (Figure~\ref{figure_RQ3_lfm}) and Industry (Figure~\ref{figure_RQ3_industry}) indicate that, in the initial stages, LLMOnline performs similarly to A-iALP. However, as the environment steps increase, A-iALP demonstrates superior performance. This can be attributed to the capacity of iALP to enable an effective exploration during the initial stage, while A-iALP refines its recommendations through continuous interaction with real-time simulated user feedback. In contrast, LLMOnline may be facing limitations as training progresses, with the model's reliance on LLM choices potentially diminishing its ability to adapt to preferences directly from the environment. Another key consideration is the time cost of online adaptation with LLMs. LLMOnline requires significant additional time for processing and generating recommendations through prompting, which can be less efficient in an online setting. In contrast, A-iALP maintains a time cost comparable to traditional RL methods, thanks to its independent pre-training phase. 

\begin{figure}[!t]
    \centering
    \begin{subfigure}[c]{0.47\textwidth}
        \centering
        \begin{minipage}[c]{0.48\textwidth}
        \centering
            \includegraphics[trim={0 20 0 0}, clip, scale=0.37]{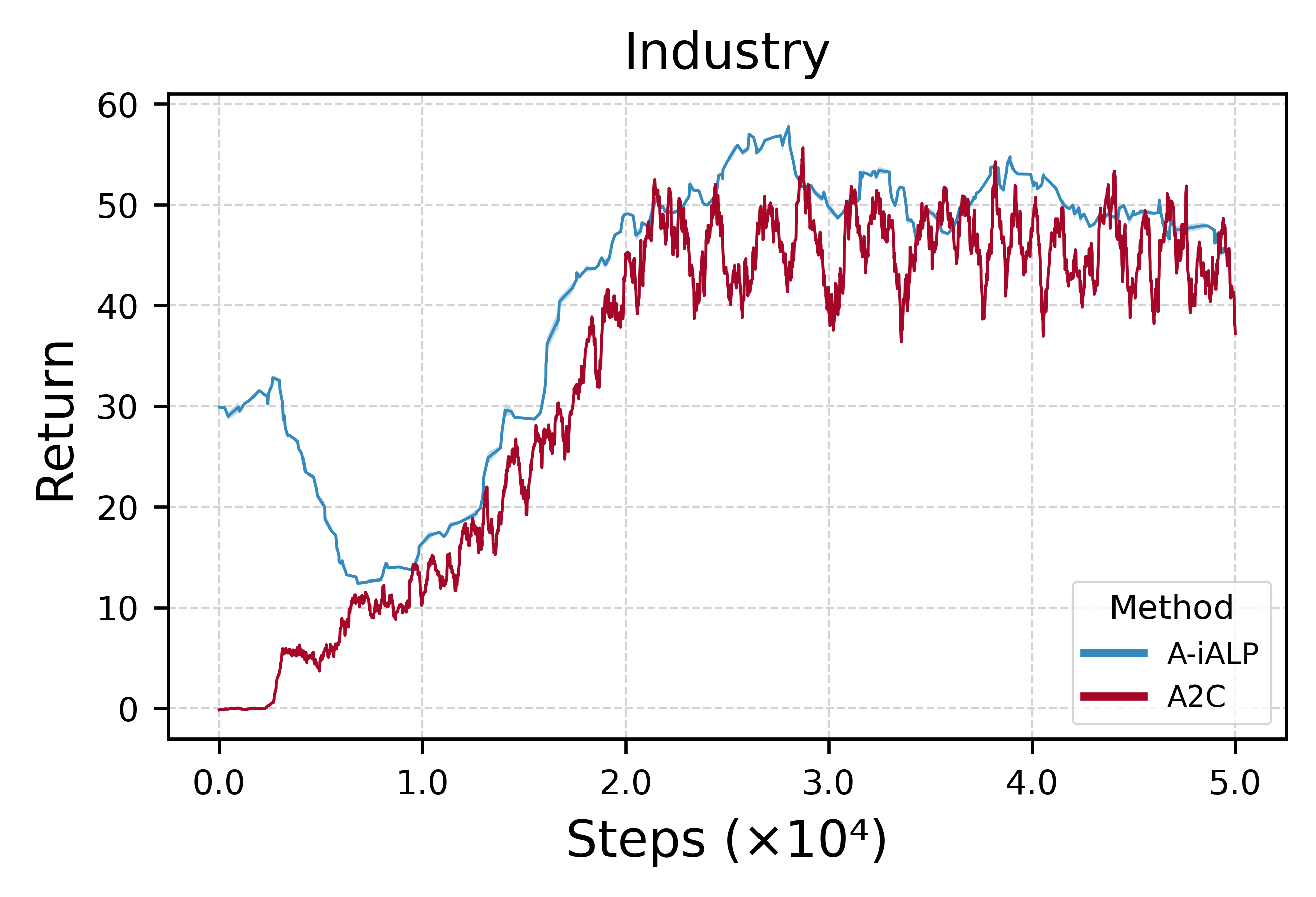}
            \label{figure1_b}
        \end{minipage}
        \hspace{0.07mm}
        \begin{minipage}[c]{0.48\textwidth}
        \centering
            \includegraphics[trim={0 20 0 0}, clip, scale=0.37]{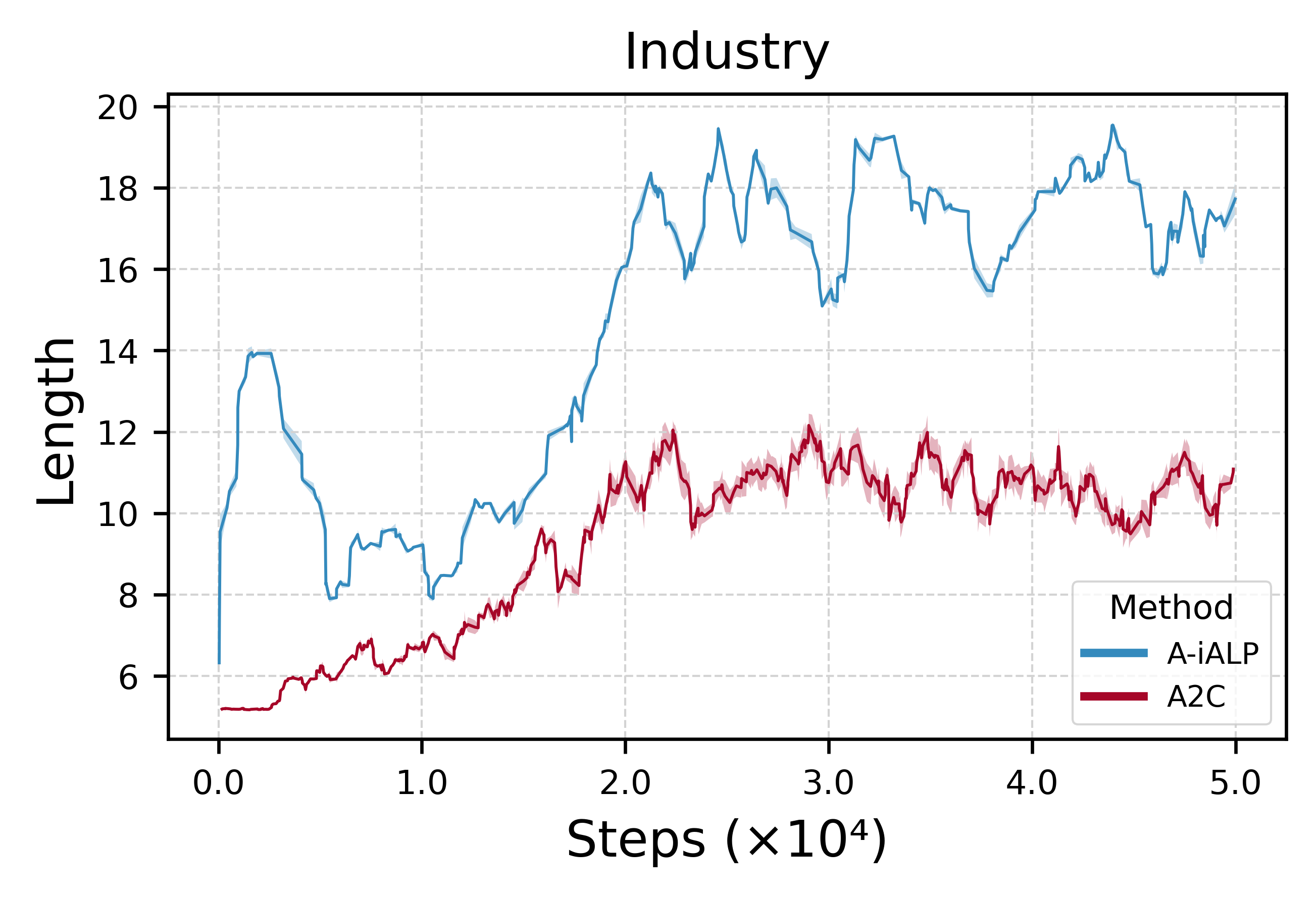}
            \label{figure1_b}
        \end{minipage}
        \vspace{-0.5cm}
        \caption{$\epsilon$-greedy}
    \end{subfigure}
    \hspace{0.07mm}
    \begin{subfigure}[c]{0.47\textwidth}
        \centering
        \begin{minipage}[c]{0.48\textwidth}
        \centering
            \includegraphics[trim={0 20 0 20}, clip, scale=0.37]{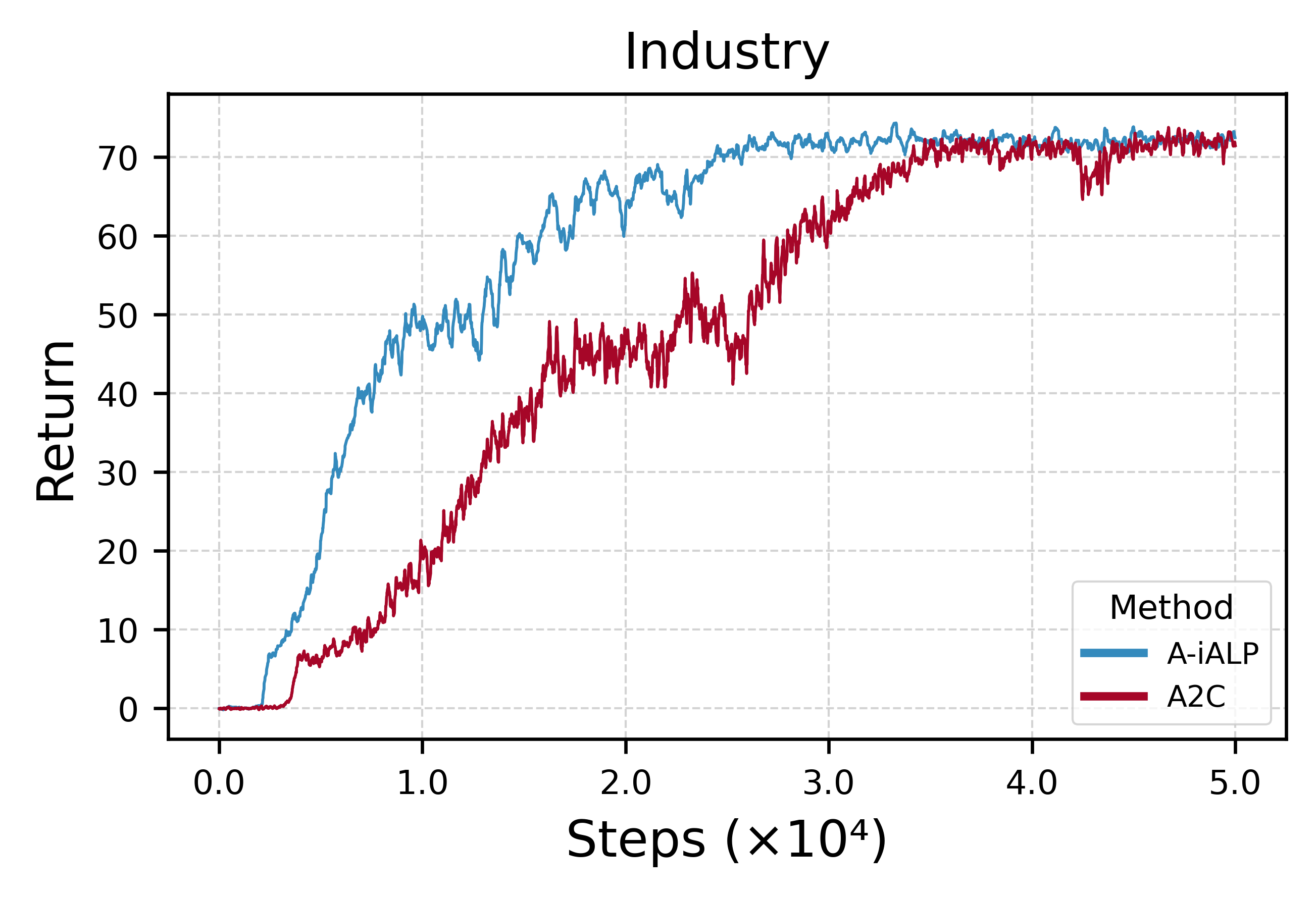}
            \label{figure1_b}
        \end{minipage}
        \hspace{0.07mm}
        \begin{minipage}[c]{0.48\textwidth}
        \centering
            \includegraphics[trim={0 20 0 20}, clip, scale=0.37]{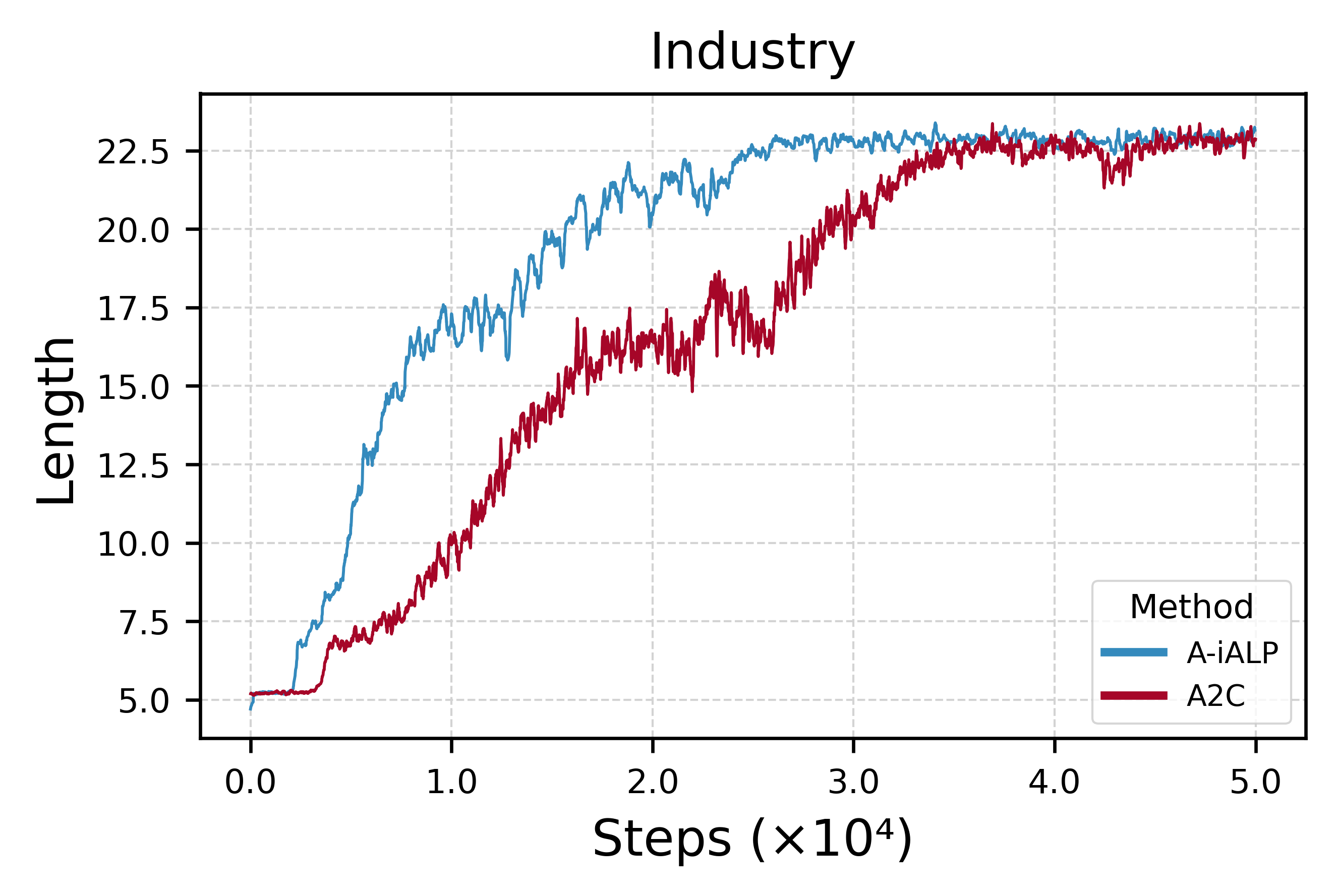}
            \label{figure1_b}
        \end{minipage}
        \vspace{-0.5cm}
        \caption{Categorical sample}
    \end{subfigure}

    \begin{subfigure}[c]{0.47\textwidth}
        \centering
        \begin{minipage}[c]{0.48\textwidth}
        \centering
            \includegraphics[trim={0 0 0 20}, clip, scale=0.37]{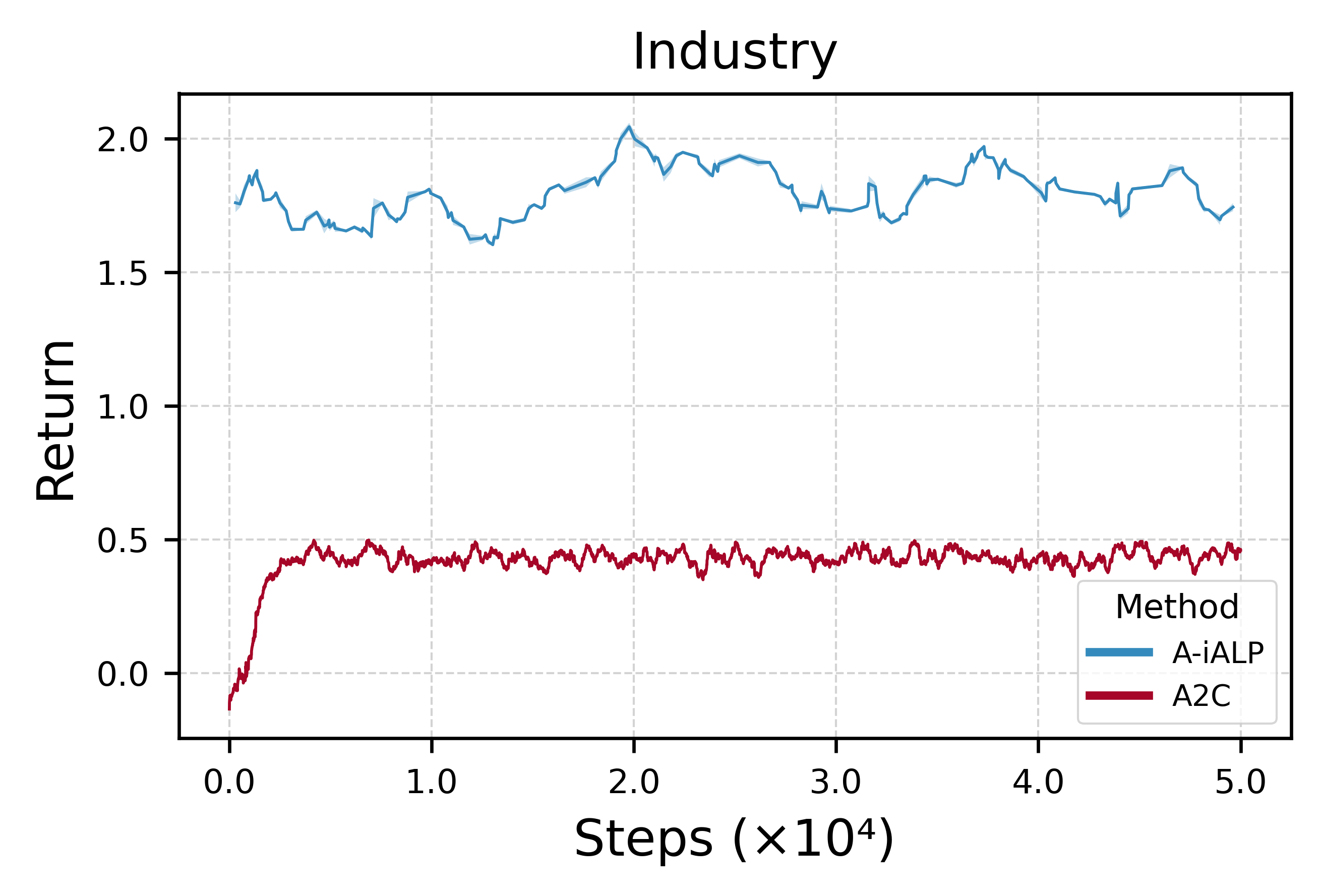}
            \label{figure1_a}
        \end{minipage}
        \hspace{0.07mm}
        \begin{minipage}[c]{0.48\textwidth}
        \centering
            \includegraphics[trim={0 0 0 20}, clip, scale=0.37]{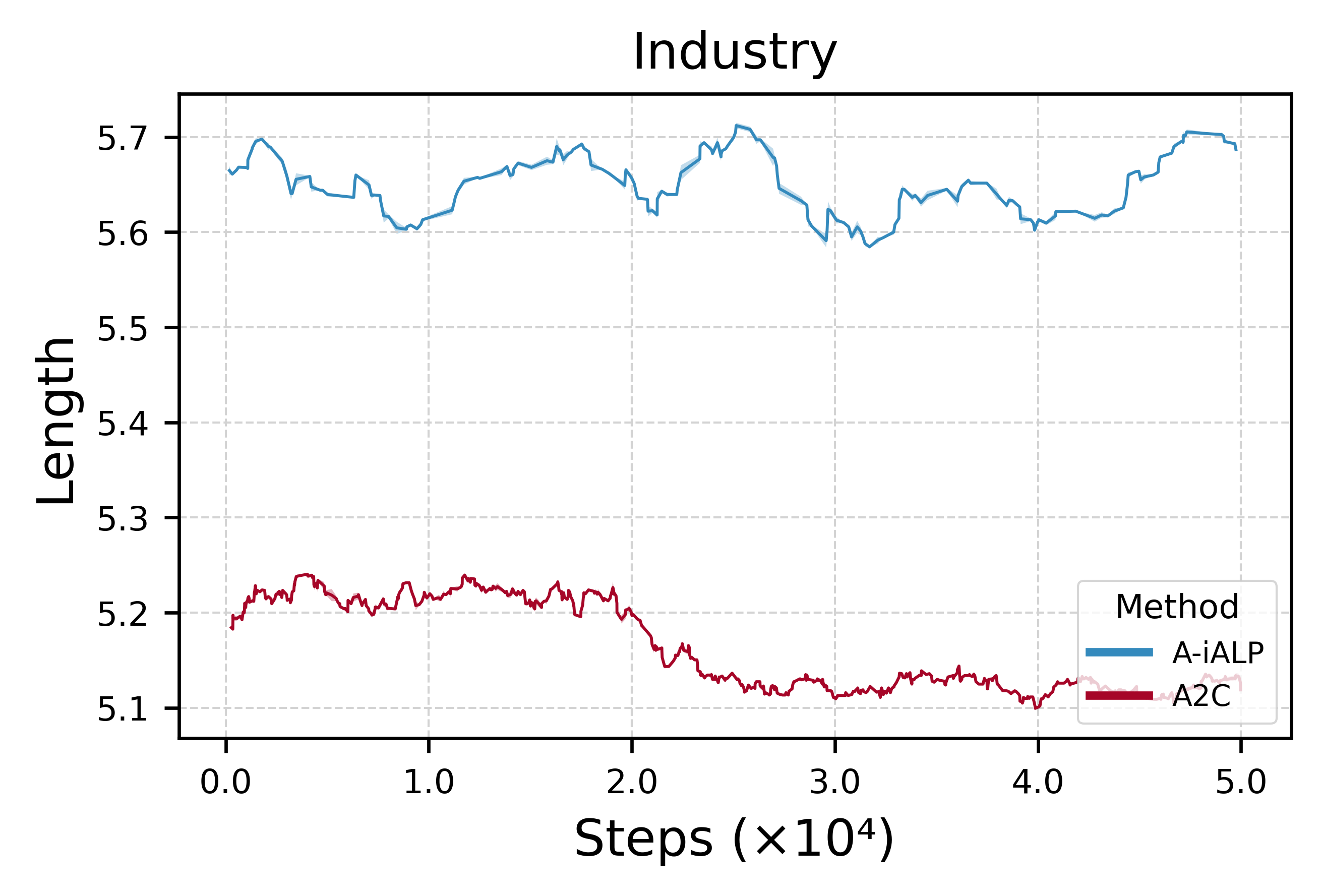}
            \label{figure1_b}
        \end{minipage}
        \vspace{-0.5cm}
        \caption{Greedy}
    \end{subfigure}
    \caption{ Performance comparison %of return (left) and length (right) 
    between A-iALP and the baseline under various exploration strategies on Industry environment.%: (a) $\epsilon$-greedy (b) Categorical sample and (c) Greedy.
    }
    \label{figure_RQ4_industry}
\end{figure}
\subsection{Effect of Exploration Strategy (RQ4)}
\begin{figure}[!t]
    \centering
    \begin{subfigure}[c]{0.47\textwidth}
        \centering
        \begin{minipage}[c]{0.48\textwidth}
        \centering
            \includegraphics[trim={0 20 0 0}, clip, scale=0.37]{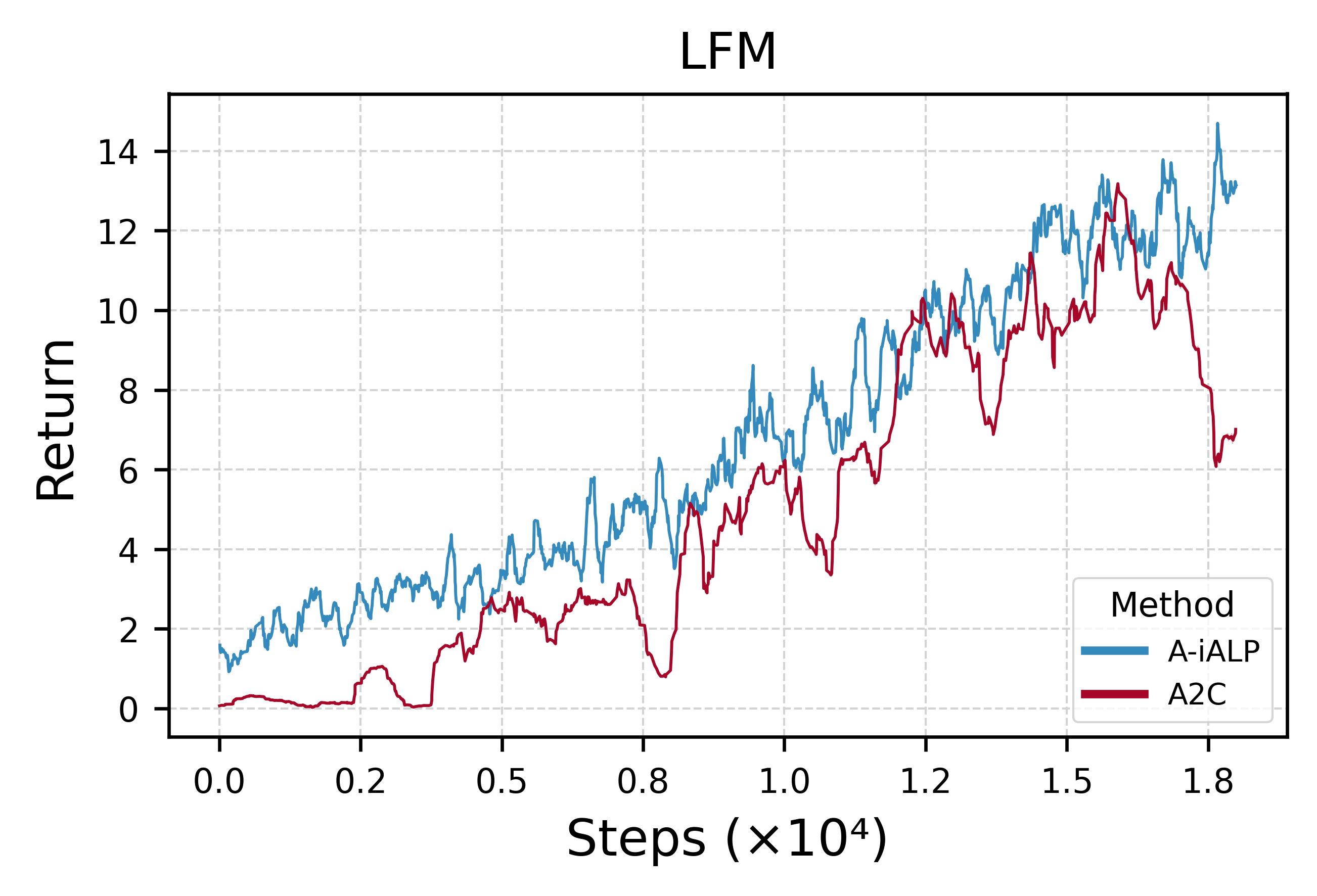}
            \label{figure1_b}
        \end{minipage}
        \hspace{0.07mm}
        \begin{minipage}[c]{0.48\textwidth}
        \centering
            \includegraphics[trim={0 20 0 0}, clip, scale=0.37]{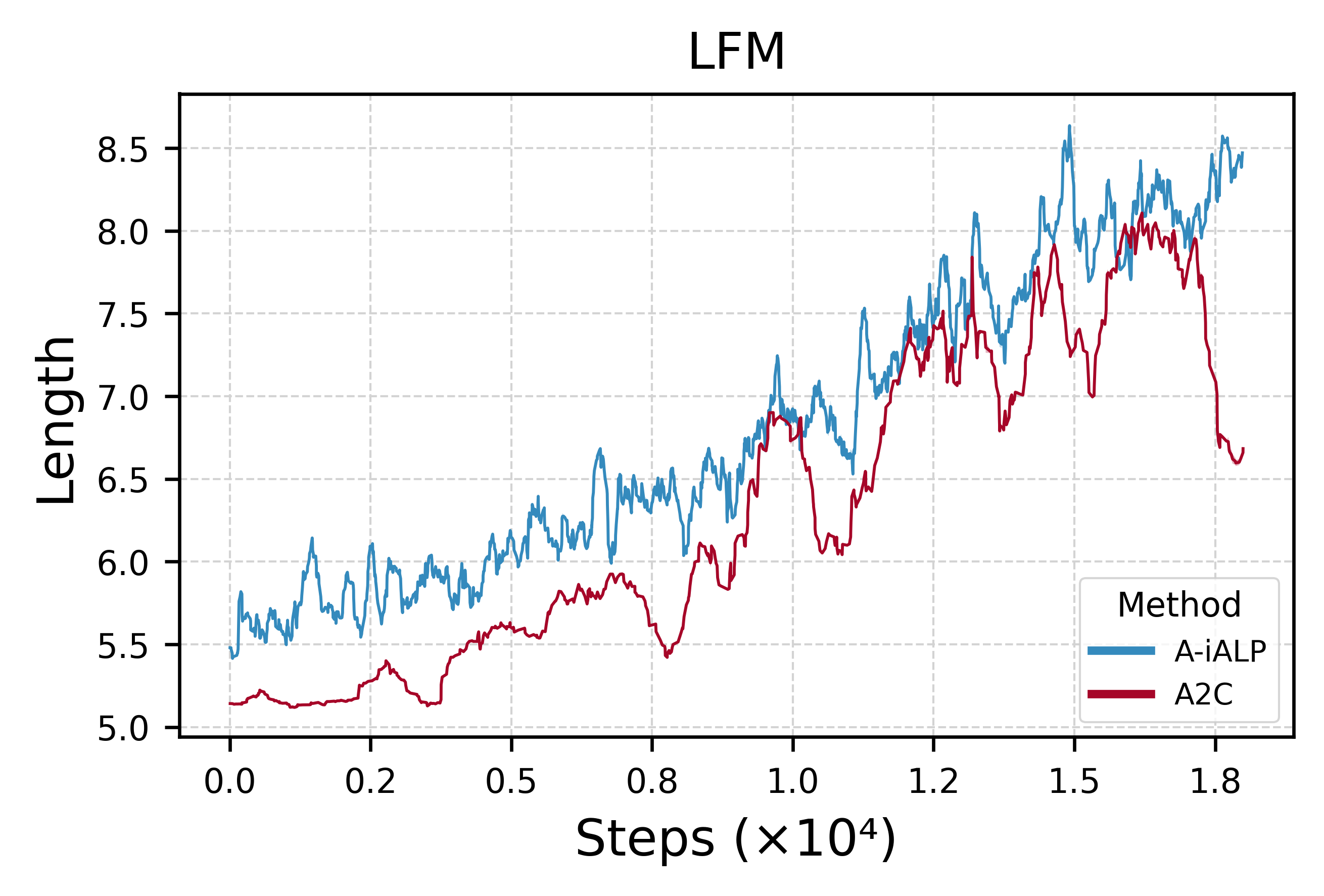}
            \label{figure1_b}
        \end{minipage}
        \vspace{-0.5cm}
        \caption{$\epsilon$-greedy}
    \end{subfigure}
    \hspace{0.07mm}
    \begin{subfigure}[c]{0.47\textwidth}
        \centering
        \begin{minipage}[c]{0.48\textwidth}
        \centering
            \includegraphics[trim={0 20 0 20}, clip, scale=0.37]{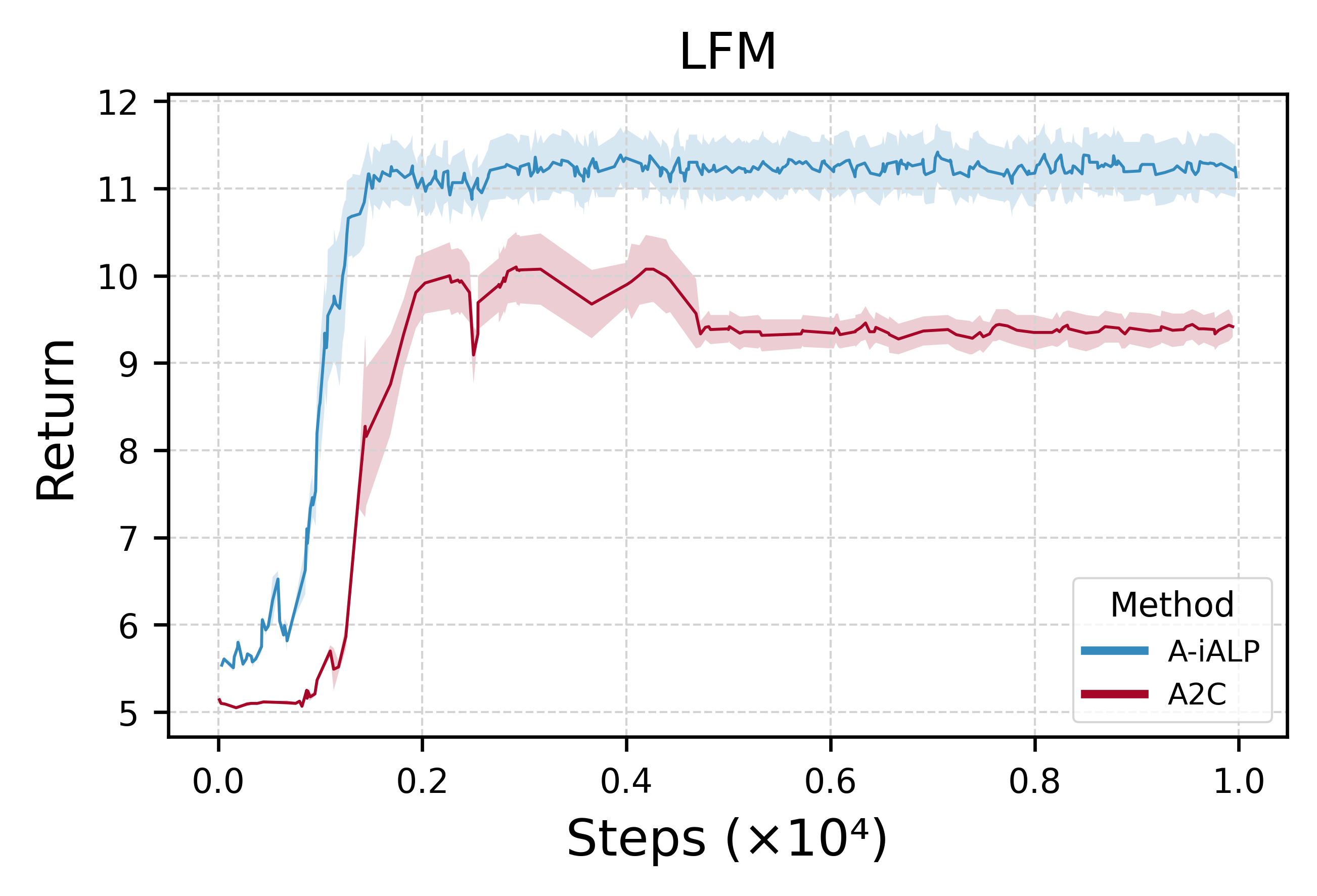}
            \label{figure1_b}
        \end{minipage}
        \hspace{0.07mm}
        \begin{minipage}[c]{0.48\textwidth}
        \centering
            \includegraphics[trim={0 20 0 20}, clip, scale=0.37]{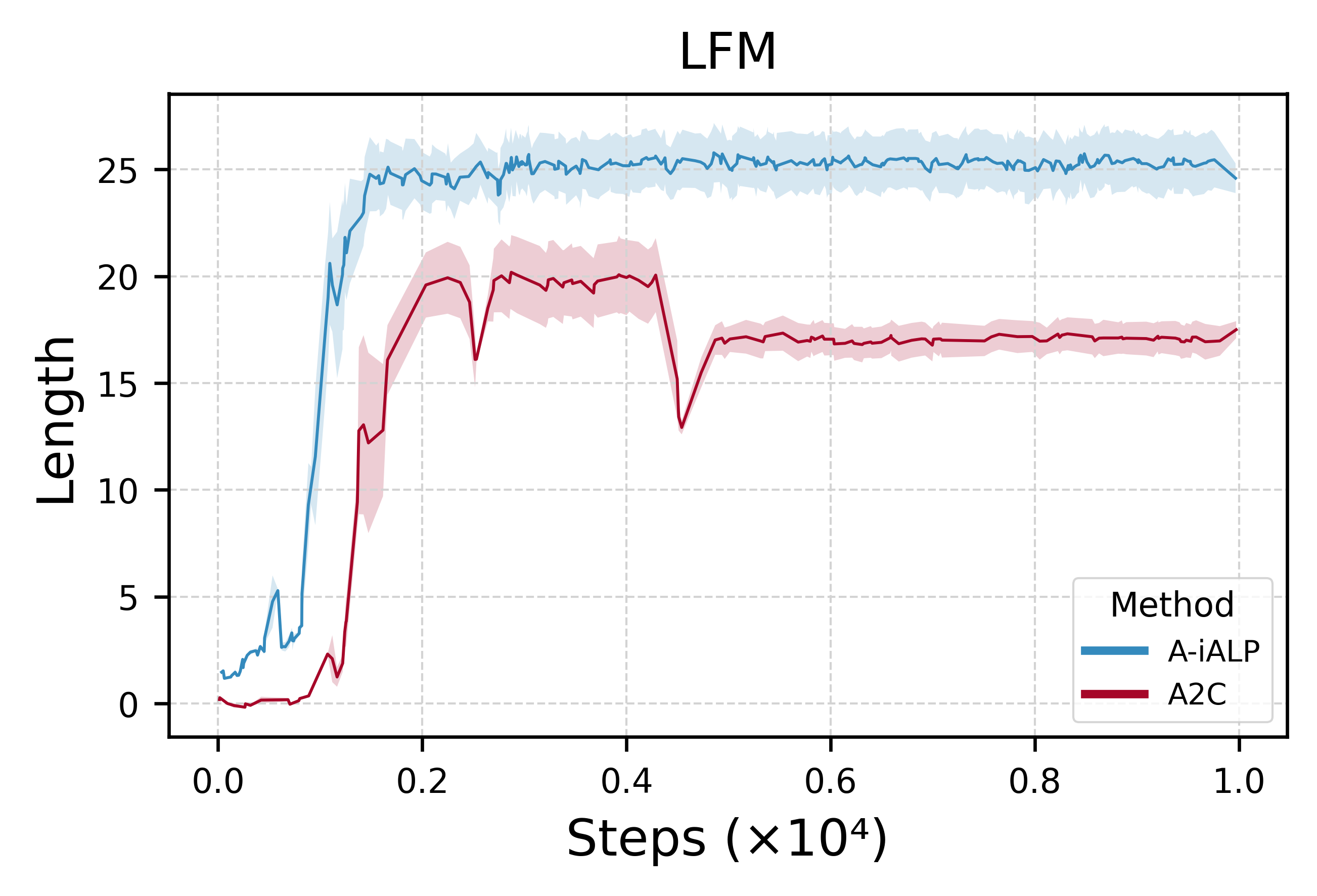}
            \label{figure1_b}
        \end{minipage}
        \vspace{-0.5cm}
        \caption{Categorical sample}
    \end{subfigure}

    \begin{subfigure}[c]{0.47\textwidth}
        \centering
        \begin{minipage}[c]{0.48\textwidth}
        \centering
            \includegraphics[trim={0 0 0 20}, clip, scale=0.37]{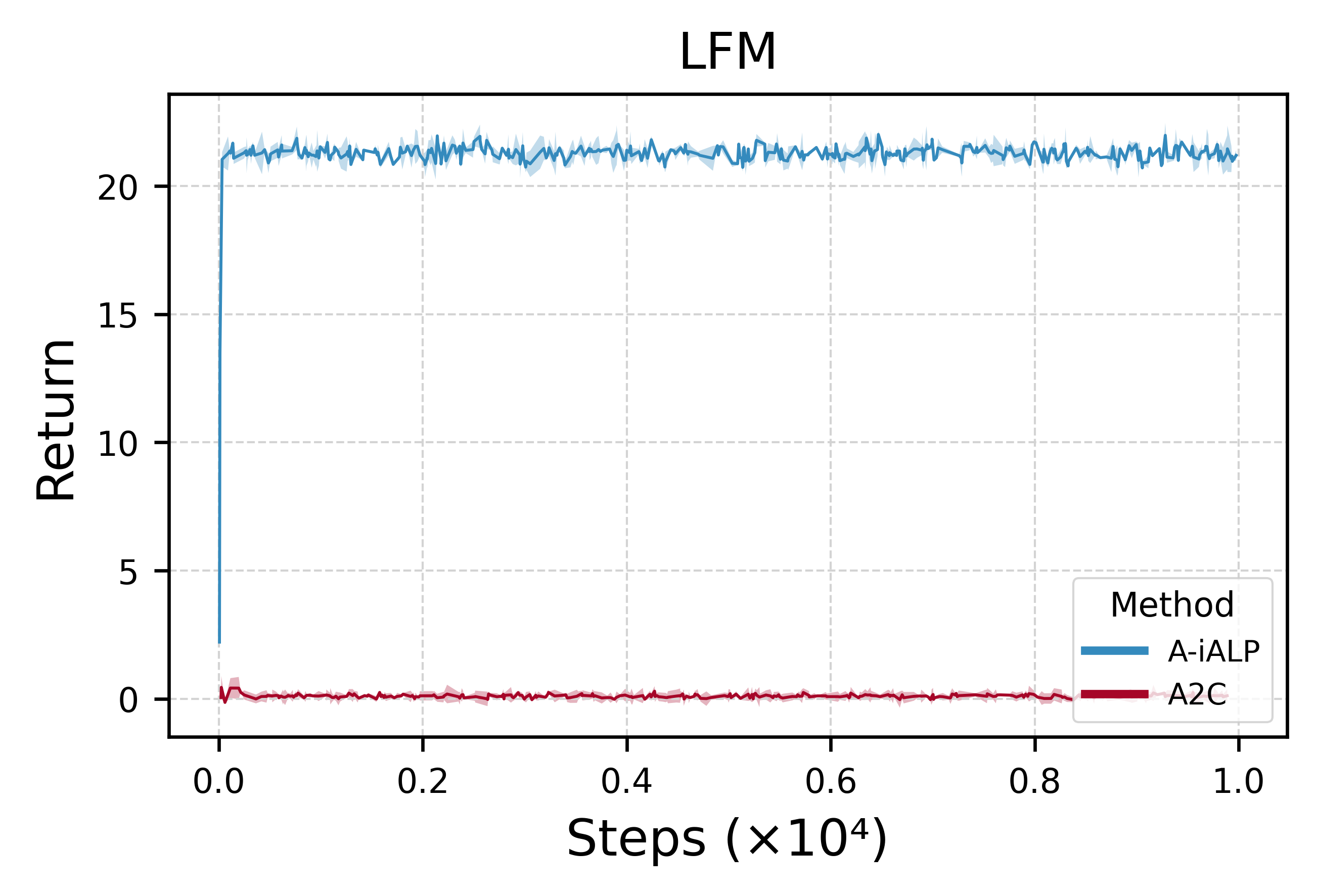}
            \label{figure1_a}
        \end{minipage}
        \hspace{0.07mm}
        \begin{minipage}[c]{0.48\textwidth}
        \centering
            \includegraphics[trim={0 0 0 20}, clip, scale=0.37]{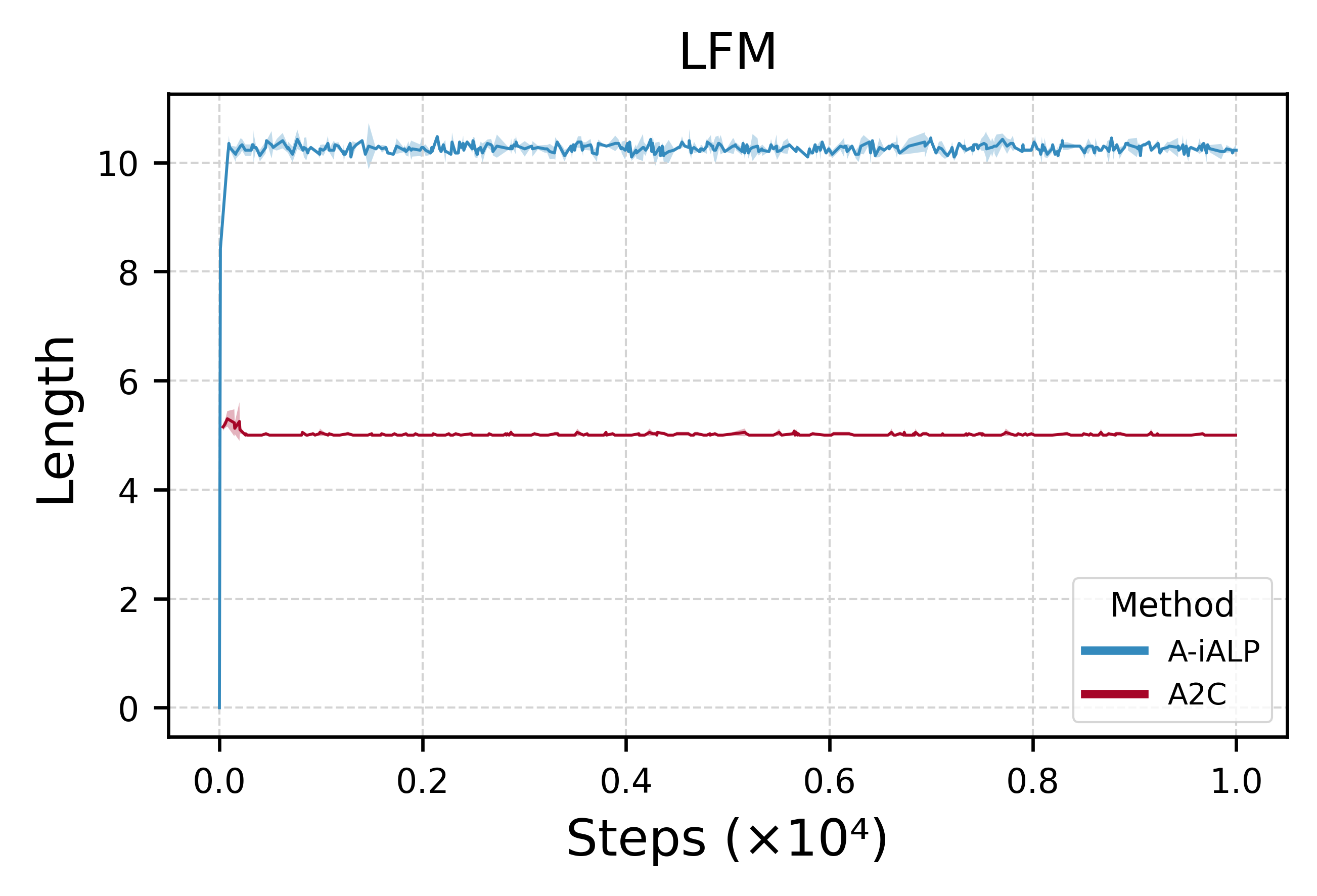}
            \label{figure1_b}
        \end{minipage}
        \vspace{-0.5cm}
        \caption{Greedy}
    \end{subfigure}
    
    \caption{ Performance comparison 
    between A-iALP and the baseline under various exploration strategies on LFM environment.
    }
    \label{figure_RQ4_lfm}
\end{figure}

Finally, we examine the impact of exploration-exploitation~\cite{kuleshov2014algorithms} strategies on A-iALP, including: (a) $\epsilon$-greedy~\cite{guo2020deep}, which explores by a random action with probability $\epsilon$ and the greedy action with probability $1-\epsilon$;  
(b) categorical sampling~\cite{liu2023exploration}, which samples actions based on the probability distribution; 
(c) greedy selection, which is a baseline that exploits the action with the highest estimation without exploration and exploitation balance.
Using LFM as a reference (Figure~\ref{figure_RQ4_lfm}), A-iALP consistently outperforms the baseline A2C method across all exploration strategies. This demonstrates the effectiveness of integrating preferences distilled from LLMs at the pre-training phase, which provides a robust foundation for more efficient exploration. Specifically, with the $\epsilon$-greedy method, A-iALP shows slightly better performance than A2C in the early training stages, with A-iALP achieving stable growth later, while A2C experiences significant fluctuations. In categorical sampling, both show stable growth, but A-iALP converges faster. Notably, under the greedy exploitation method, both maintain almost constant performance, but A-iALP's return and sequence length far exceed those of A2C. Similar results are shown in Figure~\ref{figure_RQ4_industry} for the Industry scenario. These observations demonstrate A-iALP's advantages in convergence speed and stability, leading to greater long-term gains for users.

\section{Conclusion and Future Work}
We addressed the challenges of offline and online distribution shift and the lack of data exploration in RL based recommender systems, which hinders online deplyment of RL based systems. Our investigation focused on utilizing LLMs to pre-train the RL policy, enhancing both initial user gains and long-term effectiveness in recommendation scenarios. By distilling user preferences from LLMs, we created an offline user feedback to train our RL policy, which was then adapted online using our adaptive (A-iALP) methods. Our experiments on three simulated environments demonstrated that A-iALP enhances the quality of initial recommendations while maintaining stability over time. Future research directions include expanding the pre-training dataset with more diverse user interactions and integrating advanced reward mechanisms to further optimize user satisfaction and engagement.

\bibliographystyle{ACM-Reference-Format}
\bibliography{sample-base}

\end{document}